\documentclass[fleqn,usenatbib]{mnras}

\usepackage{newtxtext,newtxmath}
\usepackage{subfigure}
\usepackage{comment}

\usepackage[T1]{fontenc}

\DeclareRobustCommand{\VAN}[3]{#2}
\let\VANthebibliography\thebibliography
\def\thebibliography{\DeclareRobustCommand{\VAN}[3]{##3}\VANthebibliography}

\usepackage{graphicx}	% Including figure files
\usepackage{amsmath}	% Advanced maths commands
\title[Studying galaxy evolution with different AGN torus models]{A comparative study of galaxy evolution with four different active galactic nucleus torus models and two different host geometries}

\author[M. Papadopoulos et al.]{
Michail Papadopoulos,$^{1}$\thanks{E-mail: mp202201@students.euc.ac.cy}
Vicky Papadopoulou Lesta,$^{1}$
Ioannis Michos,$^{1}$
Duncan Farrah$^{2,3}$ and
\newauthor
Andreas Efstathiou$^{1}$
\\
$^{1}$School of Sciences, European University Cyprus, Diogenes street, Engomi, 1516 Nicosia, Cyprus\\
$^{2}$Department of Physics and Astronomy, University of Hawaii, 2505 Correa Road, Honolulu, HI 96822, USA\\
$^{3}$Institute for Astronomy, University of Hawaii, 2680 Woodlawn Drive, Honolulu, HI 96822, USA
}

\date{Accepted XXX. Received YYY; in original form ZZZ}

\pubyear{2015}

\begin{document}
\label{firstpage}
\pagerange{\pageref{firstpage}--\pageref{lastpage}}
\maketitle

\begin{abstract}
Estimating physical quantities such as the star formation rate, stellar mass and active galactic nucleus (AGN) fraction of galaxies is a key step in understanding galaxy formation and evolution. In order to estimate the uncertainties in the predicted values for these quantities, in this paper we explore the impact of adopting four different AGN torus models in fitting the multi-wavelength spectral energy distributions (SED) of galaxies. We also explore the impact of adopting two different geometries for the host, a spheroidal geometry, more appropriate for late-stage mergers, and a disc geometry, more appropriate for galaxies forming stars with secular processes. We use optical to submillimeter photometry from the Herschel Extragalactic Legacy Project (HELP) and utilize a Markov chain Monte Carlo SED-fitting code. We use exclusively radiative transfer models for the AGN torus as well as for the starburst and host galaxy. We concentrate on a sample of 200 galaxies at $z\approx 2$, selected in the ELAIS-N1 field. All galaxies have a detection at 250$\mu m$ which ensures the presence of a starburst. We find that the stellar mass and star formation rate of the galaxies can be robustly estimated by the SED fitting but the AGN fraction depends very much on the adopted torus model. We also find that the vast majority of the galaxies in our sample are better fitted by a spheroidal geometry and lie above the main sequence.  Our method predicts systematically higher SFR and lower stellar mass than the popular energy balance method CIGALE.
\end{abstract}


%%%%%%%%%%%%%%%%%%%%%%%%%%%%%%%%%%%%%%%%%%%%%%%%%%

%%%%%%%%%%%%%%%%% BODY OF PAPER %%%%%%%%%%%%%%%%%%

\section{Introduction}

The study of galaxy evolution requires good estimates of the main physical quantities of large samples of galaxies throughout the history of the Universe. What we are mainly interested in is quantities such as stellar mass (SM), star formation rate (SFR), active galactic nucleus (AGN) fraction, etc. 
The spectral energy distribution (SED) of a galaxy contains information regarding the galactic components and the physical phenomena within galaxies. Due to the intrinsic variety of these processes, as well as the propagation of light through the interstellar medium (ISM), it is known that their footprint shows in different parts of the SED \citep[e.g.,][]{walcher2011, perez2021, pacifici2023}. Certainly, in order to understand the whole complexity of a galaxy and constrain its physical properties, high quality multi-wavelength data are required as well as methods of fitting the data with models of the emission of galaxies.

In the last two decades significant progress has been made in carrying out large surveys that cover the infrared to submillimetre part of the spectrum 
\citep{lonsdale2003,murakami2007,eales2010,elbaz2011,lutz2011,oliver2012,shirley2019,shirley2021,lacy2021}.
Nevertheless, most surveys focus on narrow parts of the spectrum and all-sky surveys still lack on depth and resolution, making multi-wavelength data from observations alone scarce and when available cross matching of observations from various surveys is necessary. An attempt towards that direction was that of the Herschel Extragalactic Legacy Project (HELP) \citep{shirley2019,shirley2021}, which made a significant contribution in the field by creating a database of homogeneously calibrated multi-wavelength catalogues of $\sim 170$ million galaxies, covering $\sim 1300\ deg^2$ of the Herschel Space Observatory \citep{pilbratt2010} survey fields. Such multi-wavelength catalogues, from local to intermediate redshift, are essential for studying galaxy evolution.

The most straightforward and established way of utilizing multi-wavelength observations in order to extract the physical properties of galaxies is that of SED fitting. With the use of physical models for the various galactic components, one is able to create synthetic SEDs and fit the data at hand. SED fitting can be done with radiative transfer models (e.g., with CYprus models for Galaxies and their NUclear Spectra or CYGNUS \citealt{farrah2003, efstathiou2021,efstathiou2022}, GRASIL \citealt{silva1998,vega2008,lofaro2015}) or with energy-balance methods (e.g., Code Investigating GALaxy Emission or CIGALE \citealt{noll2009,boquien2019}, Multi-wavelength Analysis of Galaxy Physical Properties or MAGPHYS \citealt{dacunha2008}, while there are also hybrid methods utilizing parts of both techniques (e.g., AGNfitter \citealt{rivera2016}). For recent reviews of the various SED fitting methods and  models for galactic components used in current literature and a discussion of their merits see \cite{walcher2011, perez2021, pacifici2023}.

Most large scale surveys use some form of SED fitting to estimate galaxy properties \citep[e.g.,][]{malek2018,driver2018}. However, SED fitting is a developing field of research; with no consensus regarding the fitting techniques and the models for the galactic components.  When it comes to applications, most studies use a very limited template set, sometimes as few as a handful of empirical templates, while there are rarely tests against other models.

The comparison between fitting techniques (e.g., radiative transfer vs energy-balance) is even more rare. \cite{efstathiou2021} fitted a hyperluminous obscured quasar at $z \sim 4.3$ with both CIGALE and CYGNUS, showcasing a remarkable agreement considering the limited photometry and the inherent differences of the two approaches. A more extended comparison was performed in \cite{gao2021}, where the authors compared the properties of $526$ HLIRGs derived from fitting of the SEDs with CIGALE and the Spectral energy distribution Analysis Through Markov Chains (SATMC) code \citep{johnson2013} using the CYGNUS models. The HLIRGs were fitted with two different AGN torus models with CIGALE and three different AGN torus models with CYGNUS. There appear to be no significant systematic offsets in the parameters (e.g., IR luminosity, stellar mass, SFRs, etc.), but indications of quenching of star formation by AGN are much more evident when using CYGNUS rather than CIGALE. Finally in \cite{varnava24}, a comparison of results obtained with CIGALE and SMART (Spectral energy distribution Markov chain Analysis with Radiative Transfer Models), a new SED fitting code that fits exclusively with radiative transfer models, is performed. The latter work compared the results for the HERUS sample, where the CIGALE results were taken from \cite{paspaliaris2021}. They find general agreement for the physical parameters, except that CIGALE considerably underestimates the AGN fraction, possibly because it does not consider the anisotropy correction in the AGN luminosity. As SED fitting is bound to become more and more utilized, due to deeper and more extensive extragalactic surveys producing new data, understanding the uncertainties from the fitting process itself is imperative.

In this paper, we study a sample of galaxies extracted from the HELP database in the ELAIS N1 field. While more than $50,000$ galaxies in the specific field have been fitted using CIGALE in \cite{malek2018}, in this paper we use radiative transfer methods, in particular CYGNUS \footnote{\url{https://arc.euc.ac.cy/cygnus-project-arc/}}. We plan to make a more thorough comparison between CIGALE and CYGNUS in future work, yet a brief comparison is also performed here, as $\sim 97\%$ of the galaxies in our sample were also fitted by \cite{malek2018}. Here, we fit the multi-wavelength data of a sample of 200 galaxies at $z\approx 2$ with multi-component radiative transfer models and discuss the properties of the galaxies. Our selection allows us to study galaxies which are analogous to local LIRGs/ULIRGs at $z=2$. One particular novelty of the approach that we follow is that we fit the SEDs of galaxies exclusively with radiative transfer models, eight combinations in total, employing four different models for the AGN torus \citep{efstathiou1995,fritz2006,stalevski2012,stalevski2016,siebenmorgen2015} and two models for the geometry of the host, spheroidal \citep{efstathiou2021} and disc (Efstathiou 2024, in preparation). In all eight combinations of models we include the starburst model of \cite{efstathiou2000} as updated by \cite{efstathiou2009}.

This paper is organized as follows: Section \ref{sec_data} describes the selection process and the observational data used in the analysis. Section \ref{sec_fitting}  describes the SED fitting method and the radiative transfer models that we use. In Section \ref{sec_best_fit} we show how we evaluate the quality of the fits. In Section \ref{sec_results} we describe our results and in section \ref{sec_discussion} we present our discussion and conclusions. Throughout this work, we assume $H_0 = 70$\,km\,s$^{-1}$\,Mpc$^{-1}$, $\Omega = 1$, and $\Omega_{\Lambda} = 0.7$.

\section{Data}\label{sec_data}

In this work we are exclusively dealing with data from the public database of the HELP project \citep{shirley2019,shirley2021}. The catalogue of HELP is a compilation of many multi-wavevelength surveys where low-resolution long wavelength maps have been deblended using XID+ \citep{hurley2017}. Our galaxies lie in the European Large Area ISO Survey North 1 (ELAIS-N1) field, the pilot field for HELP. ELAIS N1 is a large enough area of approximately $13.51 \ deg^2$ square degrees, containing over 4 million objects in the specific database alone. The reason we chose the ELAIS N1 field is because its size allows us to study bright and rare objects. The data generally come from various infrared surveys, mostly surveys conducted with the Herschel Space Observatory and the Spitzer Space Telescope. The data for the ELAIS N1 field in particular are obtained as part of the HerMES project \citep{oliver2012}.

Regarding our selection criteria, we focus in a narrow range of redshifts $z\in[1.9,2.1]$, which is usually known as Cosmic Noon where we have the peak of the cosmic star formation history. This is therefore an ideal epoch to examine mechanisms of star formation and AGN activity.  Another reason we have chosen to concentrate on this specific redshift is that the SED fitting method we will use relies on the use of pre-computed libraries. The use of such libraries speeds up considerably the fitting process and are computed at specific redshifts in the range 0-14 in steps of $0.2$ (Efstathiou 2024, in preparation). In future work we plan to study a much larger sample of galaxies with a range of redshifts. When referring to redshift, we refer to photometric redshift as it is available in the database of HELP and has been computed in \cite{duncan2018}, using EAZY \citep{brammer2008}.

We require from all galaxies in our sample to have a $\ge 5\sigma$ detection at $250\mu m$. At first sight, the primary selection band may appear to be inappropriate, since we are studying different AGN models and the torus is known to dominate at shorter wavelengths (e.g., $3-30\mu m$). At mid-infrared bands the torus emits anisotropically, hence using such bands as primary would bias the selection process towards strong AGN or AGN with tori which are viewed face-on. On the other hand, at sub-millimeter bands, the torus emits almost isotropically and nevertheless, we need such bands to constrain star formation as well. Beyond the primary band, we require $\ge 4\sigma$ detections in at least 12 other bands (minimum of 13 bands per galaxy). As we discuss in the following section, the number of free parameters of the fitting process are $10$ or $11$ for fits with a spheroidal or disc host respectively.  Our selection of galaxies with at least 13 photometric data points ensures that we have at least $2$ degrees of freedom per galaxy. Finally, any of the observations that do not satisfy $4\sigma$ accuracy are used as upper limits rather than being discarded completely.

Moreover, we wish to have `well-spread' detections; e.g., galaxies with multiple detections in the optical and few detections in the infrared are not considered. For this purpose we use the `herschelhelp' Python module, provided in the HELP website (\url{https://github.com/H-E-L-P}). Using this module, we filter multiple detections in a single band, retaining a single filter for each band. The choice is done in a specific order of preference, with priority given to detections in the SUBARU or MegaCam filters. In this way, we make sure that there are detections across the optical, near-infrared, mid-infrared and far-infrared parts of the spectrum.

After the processing with `herschelhelp', we find 4181 galaxies with photometric redshift $z\in[1.9,2.1]$, out of which 354 have at least 13 detections. The galaxies in our sample will later be fitted with 8 combinations of radiative transfer models. We choose a sub-sample of $200$ galaxies which is sufficient for addressing the scientific aims of this work which do not require fitting of a complete sample. The choice of the range in redshift is in any case arbitrary. We choose these $200$ galaxies to be the first ones in ascending order based on redshift. In Fig.\ref{fig_lir} in the Appendix \ref{appendix} we show how our sub-sample is representative of the initial sample, in terms of LIR as estimated by \cite{malek2018} using CIGALE. Obviously, the use of the HELP database and our requirement for at least 13 bands with $4\sigma$ accuracy will bias our sample towards moderately massive, star-forming galaxies. As our main aim in this paper however, is to assess the uncertainty in the extracted physical quantities due to different model assumptions, this will not be an important limitation in this study. 

Finally, since we aim to use a variety of models for the AGN torus, it would be very interesting to have X-ray data available for our sample. \cite{alonso2011} showed that the X-ray luminosity is well correlated with the bolometric torus luminosity. This could be very insightful in order to better constrain the properties of the torus. Unfortunately, to our knowledge there are no available X-ray data for our sample in the HELP database.

\section{SED fitting}\label{sec_fitting}

To fit the SEDs of the galaxies in our sample we use the same method employed in \cite{efstathiou2022}, where the authors fit the HERschel Ultraluminous infrared galaxy Survey (HERUS) sample \citep{farrah2013} of local ultraluminous infrared galaxies (ULIRGs). The method utilizes the Markov chain Monte Carlo code SATMC \citep{johnson2013}, as well as post-processing routines for extracting the physical quantities of interest which are described in more detail in Efstathiou (2024; in preparation). One of the novel features of this approach is the use of libraries of radiative transfer models for the starburst, AGN torus and host galaxy, which considerably reduces the computational time of the fitting of the SED per galaxy. This makes the time to fit an SED with radiative transfer models comparable to that taken by widely used energy balance methods such as  CIGALE \citep{noll2009,boquien2019} or MAGPHYS \citep{dacunha2008}.

The SED fitting method of \cite{efstathiou2022} uses four types of radiative transfer models which describe the AGN torus, the starburst, the host galaxy and the polar dust components. As the data used in this paper are limited, we do not enable the polar dust component, hence we have three types of radiative transfer models.  All models for the different galaxy components are included in the CYGNUS collection of radiative transfer models. \cite{garcia2022} also tested various torus models (smooth, clumpy, two-phase torus models and clumpy disc+wind models) in a sample of 24 Seyfert (Sy) galaxies.

\begin{enumerate}
    \item For the starburst component we use the model of \cite{efstathiou2009}, a revised version of \cite{efstathiou2000}. The fact that we use a single starburst model in this analysis is motivated by the fact that there is much less variation between the starburst model we adopted and other starburst models \citep{sie07} compared with the variation between AGN torus models which assume drastically different geometry. The starburst model assumes an identical dust mixture as the host galaxy models. The weakness of the PAH features in the starburst model is due to radiative transfer effects such as dilution from the continuum, PAH destruction and optical depth effects.

    \item For the AGN torus, we use four options that describe a broad range of geometries for the obscurer:

    \begin{itemize}

        \item The AGN torus model described in \cite{efstathiou1995} and \cite{efstathiou2013}. This model assumes a smooth tapered disc geometry, where the thickness of the disc increases linearly with distance from the black hole in the inner part of the torus, but assumes a constant thickness in the outer part.

        \item The smooth AGN torus model of \cite{fritz2006}, which assumes a flared disc geometry, where the thickness of the disc increases linearly with distance from the black hole.

        \item The two-phase AGN torus model SKIRTOR of \cite{stalevski2012,stalevski2016}, which also assumes a flared disc geometry but includes dust clumps in a smooth medium.

        \item The two-phase AGN torus model of \cite{siebenmorgen2015} which as in the case of SKIRTOR includes dust clumps in a smooth medium. Unlike the other three torus models listed above, this model assumes that dust grains are fluffy and therefore have a higher emissivity in the far-infrared and submillimetre. This model is therefore expected to give the stronger contribution from the AGN in that part of the spectrum. Moreover, the model of \cite{siebenmorgen2015} does not have the opening angle of the torus as parameter, as dust covers the whole sphere around the AGN. Therefore, this model essentially combines the concepts of an AGN torus and polar dust.
    \end{itemize} We did not fit with a clumpy torus model as this type of model is currently not included as an option in our fitting method. We note, however, that recently Garcia-Bernete et al. (2024) concluded by analyzing JWST data that smooth models perform better than clumpy models in explaining the properties of six local obscured type 1.9/2 AGN.

    \item For the geometry of the host galaxy we have two options, a spheroidal host \citep{efstathiou2021} and a disc host (Efstathiou 2024, in preparation). Both models assume an age of the model galaxies of  $3.223$ Gyr which is the age of the Universe at z=2 for the adopted cosmology.

\end{enumerate}

In total, we have four options for the AGN torus and two options for the geometry of the host galaxy. To explore the impact of these models on the results, we fit the SEDs in our sample with all 8 combinations of models. Moreover, since SATMC is based on MCMC it is possible for the Markov chains to be stuck in some local extrema. Hence, for each of the $8$ combinations of models we compute 3 different runs, keeping the best fitting one. This amounts to $24$ different runs for each galaxy in our sample. The parameters and their range are shown in detail in Table \ref{tab:fiting_parameters} and here in summary:

\begin{itemize}

\item Starburst (2 parameters): starburst age, starburst optical depth (we fix starburst star formation rate e-folding time to $t_e=2\times 10^7$yr)

\item AGN torus (2 parameters): torus optical depth, torus inclination (we fix 2 more parameters for each model, see Table \ref{tab:fiting_parameters}. The ratio of outer to inner torus radius, a parameter that we fixed, can be important for the NIR-to-FIR slope, meaning that it can affect the contribution of the starburst. In order to test whether this is the case, we fitted a galaxy selected at random ($HELP\_J161123.428+535540.996$) two times for each AGN model. For the first set of fits we leave the parameter free, assuming the same range for each AGN model assumed by \cite{efstathiou2022}. For the second set of fits, we fixed the parameter to the values indicated in Table \ref{tab:fiting_parameters}. As shown in Fig.\ref{fig_torus_radial_extent} in the Appendix \ref{appendix}, we find a very small difference in SFR (negligible for the CYGNUS model and about 15-20\% for the fits with the Fritz and SKIRTOR models). This difference could quite possibly be due to the stochastic nature of the MCMC process within SAMTC, rather than the parameter itself. We therefore proceed with fixing the parameter according to Table \ref{tab:fiting_parameters}.

\item Spheroidal host (3 parameters): spheroidal optical depth, starlight intensity $\psi$, spheroidal e-folding time of the delayed exponential star formation history

\item Disc host (4 parameters): inclination, disc optical depth, starlight intensity $\psi$, disc e-folding time of the delayed exponential star formation history

\end{itemize}

Additionally, there is an extra parameter for each component ($f_{sb}$, $f_{AGN}$, $f_{sph}$, $f_{disc}$), scaling factors for the models. In general, there are 10 free parameters for the fits using the spheroidal host, while there are 11 free parameters for the fits using the disc host.

Some careful consideration is needed when fitting with the \cite{fritz2006} and \cite{siebenmorgen2015} AGN torus models. For the \cite{fritz2006} model, there is an issue with the optical depth of the torus. If we use the whole range available in the library we get solutions where the AGN contributes significantly in the optical and near-IR and as a result gives unrealistically small stellar masses as part of the solution. Therefore, we assume a range of $9.7\mu m$ optical depth of [3,10].  For the same reason, when using \cite{siebenmorgen2015}, we assume a minimum optical depth in the mid-plane of 50, a factor of 10 below the maximum. Finally, for all AGN models, we restrict the range of inclination in the range [$45-90^\circ$]. Essentially, this means that we are fitting the galaxies only with type 2 AGN. Ideally we should also fit with type 1 AGN, allowing us to identify broad line AGN in our sample, however in practice we find very few cases where a type 2 AGN is not a good solution.

Finally, we utilize post-processing routines that calculate self-consistently from the tables of \cite{bru93,bru03}, on which the radiative transfer models are based, quantities such as stellar mass, star formation rate etc. The properties that we obtain are shown in Table \ref{tab:physical_quantities}, along with the respective contributions from the various galactic components. Since we are dealing with the output of an MCMC code, some parameters may be well constrained in the parameter space, while others are not. Moreover, since the dependence of the output on the parameters is generally non-linear, the errors of the parameters are asymmetric (e.g., for some parameter $\alpha$, $\alpha^{+\sigma^+}_{-\sigma^-}$). In order to accept a parameter as `well constrained', we require that the value of the parameter is larger than $2\sigma$, where $\sigma = \max\{\sigma^+,\sigma^-\}$.

\begin{table*}
\centering
\caption{Parameters of the models used in this paper, symbols used, their assumed ranges and summary of other information about the models. The parameters that are fixed are shown in brackets. The Fritz et al. model assumes two additional parameters that define the density distribution in the radial direction ($\beta$) and azimuthal direction ($\gamma$). In this paper we assume $\beta=0$ and $\gamma=4$. The SKIRTOR model assumes two additional parameters that define the density distribution in the radial direction ($p$) and azimuthal direction ($q$). In this paper we assume $p=1$ and $q=1$. In addition the SKIRTOR library we used assumes the fraction of mass inside clumps is 97\%.}
	\label{tab:fiting_parameters}
	\begin{tabular}{llll} % four columns, alignment for each
		\hline
		Parameter &  Symbol & Range &  Comments\\
		\hline
                 &  &  & \\
{\bf CYGNUS Starburst}  &  &  & \\
               %  &  &  \\
Initial optical depth of giant molecular clouds & $\tau_V$  &  50-250  &  \cite{efstathiou2000}, \cite{efstathiou2009} \\
Starburst star formation rate e-folding time       & $\tau_{*}$  & (20 Myr)  & Incorporates \citet{bru93,bru03}  \\
Starburst age      & $t_{*}$   &  5-35 Myr &  metallicity=solar, Salpeter Initial Mass Function (IMF) \\
                  &            &  & Standard galactic dust mixture with PAHs \\
                 %  &           &         &    \\
{\bf CYGNUS Spheroidal Host}  &  &  &  \\
               %  &  &  \\
Spheroidal star formation rate e-folding time      & $\tau^s$  &  0.125-8 Gyr  & \cite{efstathiou2003}, \cite{efstathiou2021}  \\
Starlight intensity      & $\psi^s$ &  1-17 &  Incorporates \citet{bru93,bru03} \\ 
Optical depth     & $\tau_{v}^s$ & 0.1-15 &  metallicity=6.5\% of solar, Salpeter IMF\\ 
                  &            &  & Standard galactic dust mixture with PAHs \\
                 &            &  &  \\

{\bf CYGNUS Disc Host}  &  &  &  \\
              %   &  &  \\
Disc SFR e-folding time      & $\tau^d$  &  0.5$-$8 Gyr  & \cite{efstathiou2003}, Efstathiou (2024, in prep.)  \\
Starlight intensity      & $\psi^d$ &  1$-$9 &  Incorporates \cite{bru93,bru03} \\
Optical depth     & $\tau_{v}^d$ & 2$-$29 &  metallicity=6.5\% of solar, Salpeter IMF\\
Inclination       &  $\theta_d$    & 0\degr$-$90\degr &     Standard galactic dust mixture with PAHs     \\
                  &            &  &  \\
{\bf CYGNUS AGN torus}  &  &    &  \\
              %   &  &  &  \\
Torus equatorial UV optical depth   & $\tau_{uv}$  &  250-1450 &  Smooth tapered discs\\  
Torus ratio of outer to inner radius & $r_2/r_1$ &  (60) & \cite{efstathiou1995}, \cite{efstathiou2013} \\   
Torus half-opening angle  & $\theta_o$  &  (45\degr) & Standard galactic dust mixture without PAHs\\ 
Torus inclination     & $\theta_i$  &  45-90\degr &  \\ 
                 &            & \\
{\bf Fritz AGN torus}  &  &  &   \\
               %  &  &  & \\
Torus equatorial optical depth at 9.7$\mu m$  &  &  3-10 & Smooth flared discs \\  
Torus ratio of outer to inner radius &  &  (60) &  \cite{fritz2006}\\   
Torus half-opening angle  &   &  (45\degr) & Standard galactic dust mixture without PAHs\\ 
Torus inclination     &   &  45-90\degr &  \\   
                 &            & &  \\
{\bf SKIRTOR AGN torus}  &  &   &  \\
               %  &  &  &  \\
Torus equatorial optical depth at 9.7$\mu m$  &   &  3-11 & Two-phase flared discs \\  
Torus ratio of outer to inner radius &  &  (20) &  \cite{stalevski2012}, \cite{stalevski2016} \\   
Torus half-opening angle  &  &  (45\degr) &  Standard galactic dust mixture without PAHs\\ 
Torus inclination     &   &  45-90\degr &  \\ 
                 &            & &  \\
{\bf Siebenmorgen15 AGN torus}  &  &   &  \\
               %  &  &  &  \\
Cloud volume filling factor (\%)   &  &  (40)  & Two-phase anisotropic spheres \\  
Optical depth of the individual clouds &   &  (30) & \cite{siebenmorgen2015}\\
Optical depth of the disk mid-plane &   &  50-500 &  Fluffy dust mixture without PAHs\\ 
     Inclination     &   &   45-90\degr & \\   
		\hline
	\end{tabular}
\end{table*}

\begin{table}
	\centering
\caption{Derived physical quantities and the symbol used. The luminosities are integrated over 1-1000$\mu m$.}
	\label{tab:physical_quantities}
	\begin{tabular}{ll} % four columns, alignment for each
		\hline
		Physical Quantity &  Symbol \\
		\hline
                                       &                \\     
Observed AGN torus Luminosity          & $L_{AGN}^{o}$  \\  
Corrected AGN torus Luminosity         & $L_{AGN}^{c}$  \\
Starburst Luminosity                   & $L_{Sb}$ \\   
Spheroidal host Luminosity                  & $L_{host}$   \\ 
Total observed Luminosity              & $L_{Tot}^{o}$    \\  
Total corrected Luminosity             & $L_{Tot}^{c}$    \\   
Starburst SFR (averaged over SB age)   & $\dot{M}_*^{age}$   \\
Spheroidal SFR                         & $\dot{M}_*^{Sph}$    \\
Total SFR                              & $\dot{M}_{tot}$    \\ 
Starburst Stellar Mass                 & $M^{*}_{sb}$  \\
Spheroidal Stellar Mass                & $M^{*}_{sph}$  \\ 
Total Stellar Mass                     & $M^{*}_{tot}$ \\   
AGN fraction                           & $F_{AGN}$  \\ 
Anisotropy correction factor           & $A$  \\
		\hline
	\end{tabular}
\end{table}

\subsection{SED fitting quality test}\label{sec_best_fit}

In order to estimate the quality of the SED fitting, we use the $\chi^2$ statistic. If we denote with $f_{obs}$ the observations of a galaxy that we are trying to fit and with $f_{mod}$ the estimated SED derived from one of our configuration of models, $\chi^2$ is calculated to be:

\begin{equation}
    \chi^2 = \frac{\sum (f_{obs} - f_{mod})^2}{\sigma_{obs}^2+\sigma_{mod}^2} 
\end{equation}

\noindent where $\sigma_{obs}$ and $\sigma_{mod}$ are the errors of the observation and estimated flux respectively. Regarding the errors of the estimated fluxes, it is standard practice to account for a systematic error which we set to $20\%$ of each flux (this error can be as high as $30\%$, e.g., \citealt{lanz2014}). This accounts for any uncertainties in the models due to different assumptions, dust properties, metallicity etc. Concerning the observations, there are available statistical observational errors in the database of HELP $(\sigma_{stat})$, but those errors can be way too small (as low as $0.1\%$ of the observed flux). This can lead to a bias towards the fitting parameters, while also heavily penalizing our $\chi^2$ statistic. In order to overcome this, we add an additional systematic error of $15\%$ in all observations in the following way:
\begin{equation}
    \sigma_{obs}^2 = \sigma_{stat}^2 + (15\% \times observed\_flux)^2    
\end{equation}

When interested for goodness of fit testing, as in our case, it is important to account for the  degrees of freedom (dof) of the fitting process ($\nu = \# \text{ of observations} - \# \text{ of free parameters}$). Alternatively, the reduced $\chi_{\nu}^2$, a normalization of $\chi^2$ per degree of freedom defined as $\chi_{\nu}^2 = \chi^2/\nu$, is principally used. The $\chi^2$-test for goodness of fit identifies a threshold value $\chi^2_{th}$ and rejects the fit if $\chi^2_{\nu}>\chi^2_{th}$. In our work we fix this threshold value to $\chi^2_{th} = 5$, a value that corresponds to $99.9\%$ confidence interval (or better) for all galaxies in our sample, regardless of the degrees of freedom.

\begin{figure*}
\centering
\subfigure[]{\includegraphics[width=0.49\textwidth]{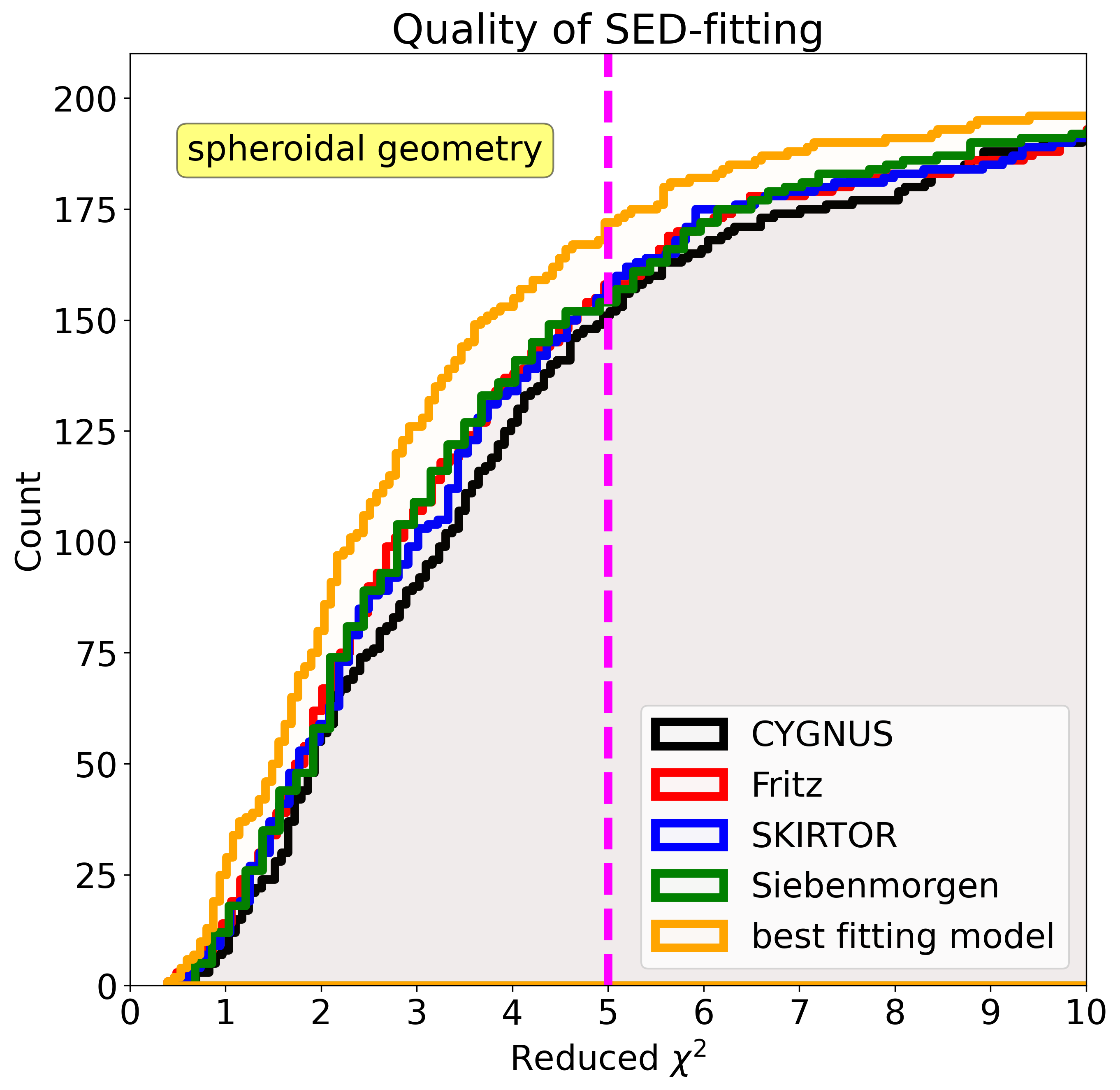}\label{chi_sph}}
\subfigure[]{\includegraphics[width=0.49\textwidth]{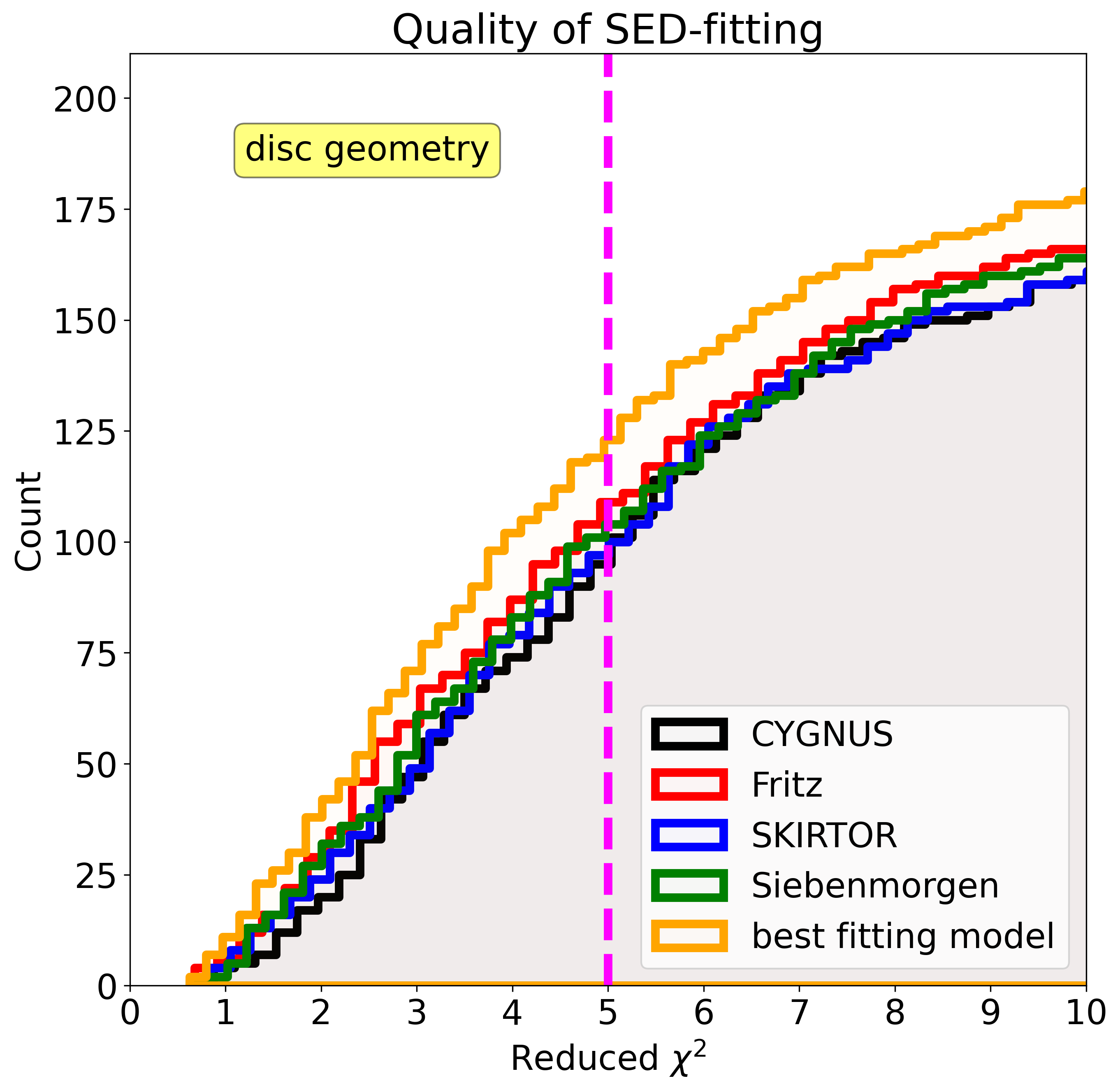}\label{chi_disc}}
\caption{Reduced $\chi^2$ for the various configurations of radiative transfer models. In (a) we show the quality of the fits for the spheroidal host and in (b) for the disc host. For the spheroidal geometry, each model fits roughly $150$ galaxies, while we can fit 170 galaxies with at least one AGN model. For the disc host, each model fits roughly $100$ galaxies, while we fit $121$ galaxies with at least one AGN model. In general, the spheroidal geometry seems to fit the galaxies better, regardless of the assumed AGN model. Moreover, any galaxy that can be fitted with a disc host can also be fitted with a spheroidal host, while the opposite is never the case.}
\label{fig_chi_square}
\end{figure*}

\subsection{Radiative transfer models versus energy balance methods}
We prefer to use radiative transfer models in this paper instead of energy balance methods for a number of reasons. First of all, the use of radiative transfer models for SED decomposition purposes has a long history, starting with \cite{robinson1989}, where the authors fitted the SEDs of 227 infrared galaxies detected by the Infrared Astronomical Satellite (IRAS). Through the years, more detailed models have been developed regarding the different galaxy components; namely the starburst component \citep[e.g.,][]{robinson1993,kru94,silva1998,efstathiou2000,tagaki2003,dop05,sie07,efstathiou2009}, the AGN torus \citep[e.g.,][]{pier1992,pier1993,granato1994,efstathiou1995,nenkova2002,dullemond2005,honig2006,fritz2006,schartmann2008,nenkova2008,stalevski2012,siebenmorgen2015,stalevski2016,honig2017} and the host galaxy \citep[e.g.,][]{efstathiou2003,efstathiou2021}.

Moreover, energy balance methods such as CIGALE \citep{noll2009,boquien2019} and MAGPHYS \citep{dacunha2008}) rely on the principle of conservation of energy and are an approximation of the complex radiative transfer process. Generally, such methods (1) assume an intrinsic spectrum of the stellar population of the galaxy, (2) assume an attenuation law \citep[e.g.,][]{charlot2000,calzetti2000} to convert the intrinsic spectrum into the observed one, (3) assume a model for dust emission \citep[e.g.,][]{draine2007,casey2012,dale2014,draine2014} and convert the difference of the intrinsic and observed spectrum to infrared and submillimeter emission. They are known to have limitations in general (e.g., an attenuation law might not be able to capture the complex geometry of some galaxies) and the main reason that they  are currently favoured over radiative transfer models is the computational cost that the latter require. It is important to note that some energy balance methods incorporate radiative transfer models for the AGN torus, e.g., in the recent versions of CIGALE, the models of \cite{fritz2006} and \cite{stalevski2016} are incorporated. Such approaches are therefore a combination of the two fitting methods. We note that CIGALE is categorized as both an energy balance and hybrid method in the categorization of \cite{perez2021}. Moreover, with the use of pre-computed libraries, the process of fitting SEDs with radiative transfer models is made considerably faster (as in e.g., \citealt{verma02,farrah02,farrah2003,farrah16,herrero2017,kankare21,efstathiou2021,efstathiou2022,varnava24}). Therefore, here we explore SED fitting exclusively with radiative transfer models. Nevertheless, as there is large overlap between our sample and the sample in \citep{malek2018} ($193$ out of our $200$ galaxies), we make a brief comparison with the CIGALE output obtained there. As part of future work, we plan to carry out a more extensive comparison between energy balance and radiative transfer methods.

\section{Results}\label{sec_results}

\begin{figure*}
\centering
Physical properties of galaxies (spheroidal geometry host)\par\smallskip
\subfigure[]{\includegraphics[width=0.49\textwidth]{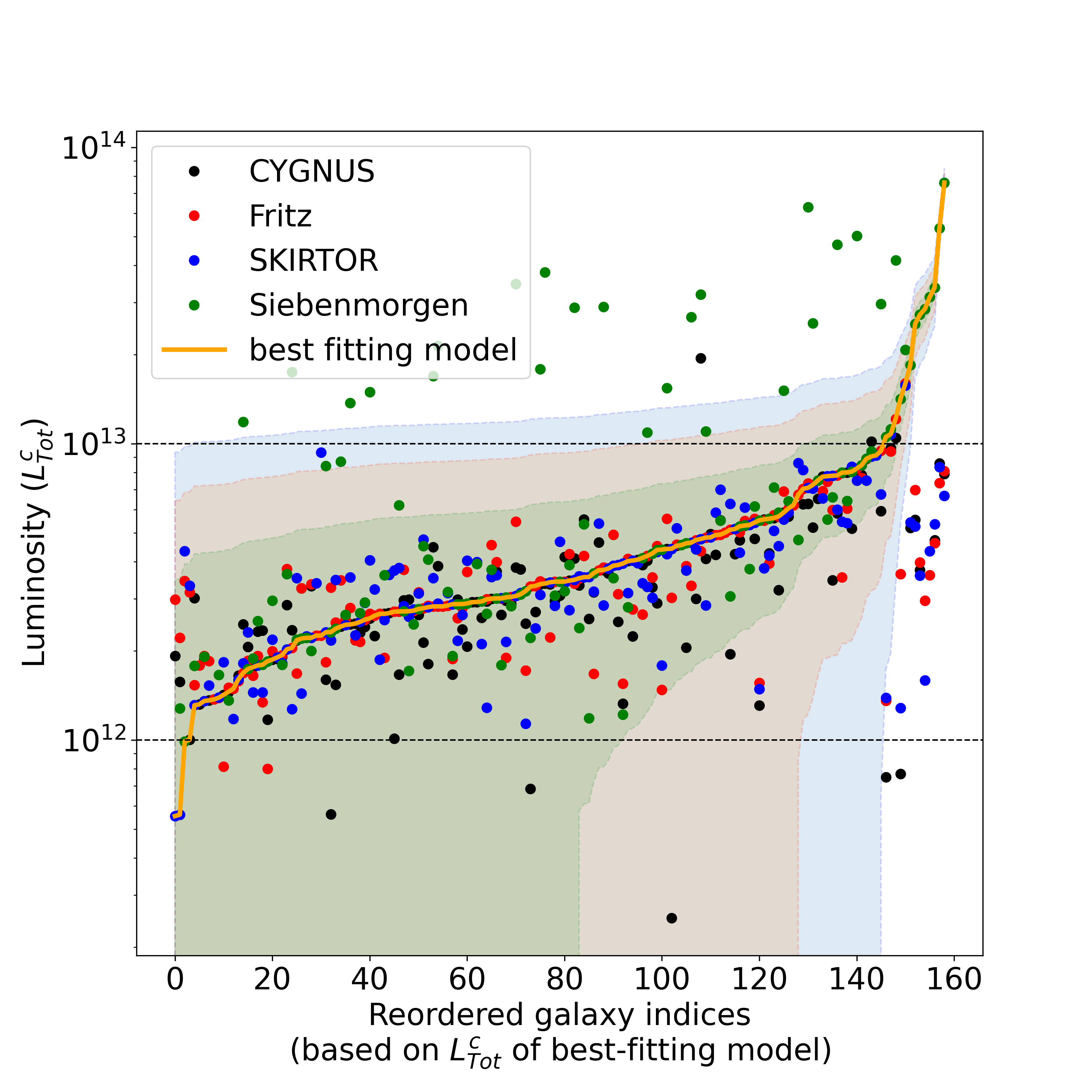}\label{lum_sph}}
\subfigure[]{\includegraphics[width=0.49\textwidth]{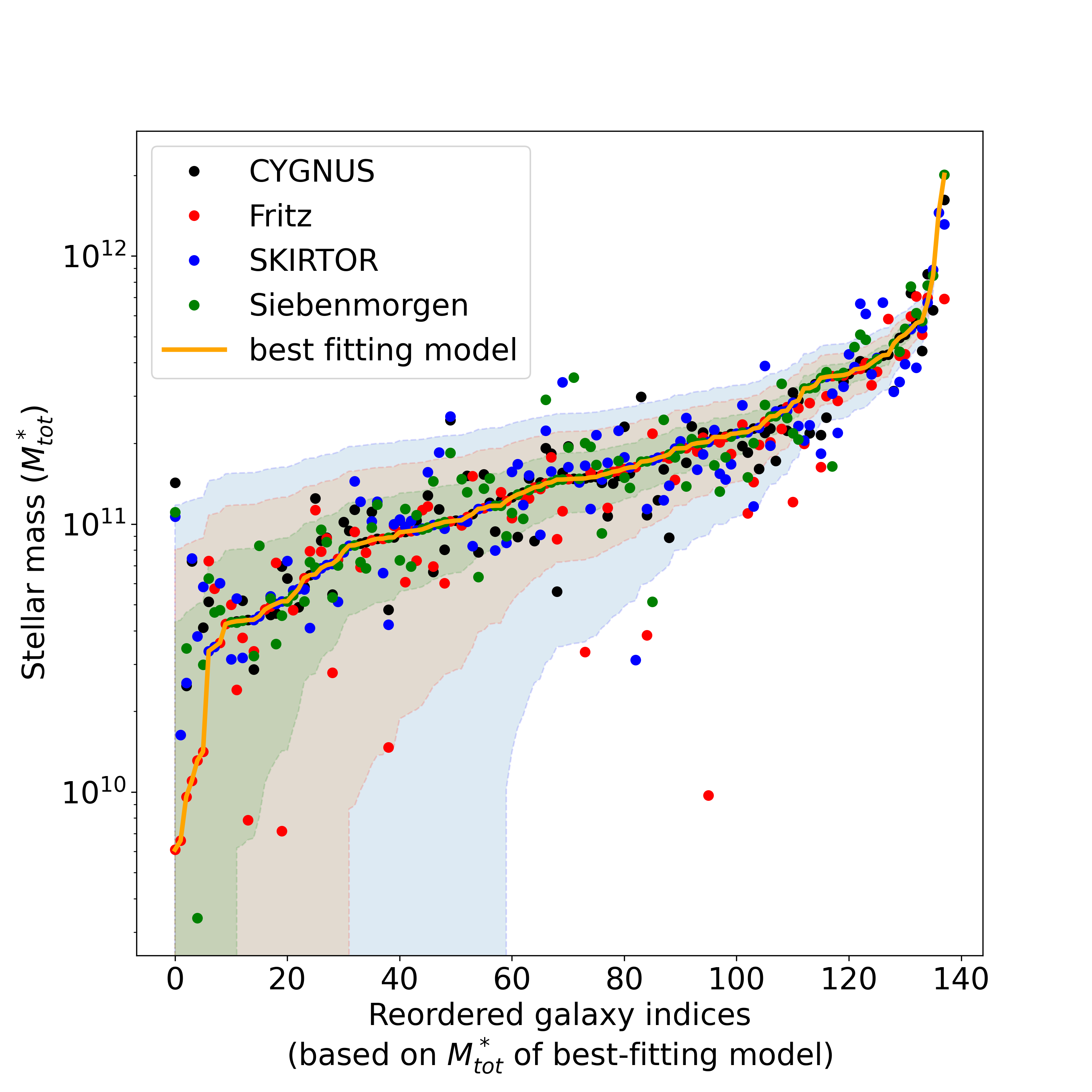}\label{sm_sph}}
\subfigure[]{\includegraphics[width=0.49\textwidth]{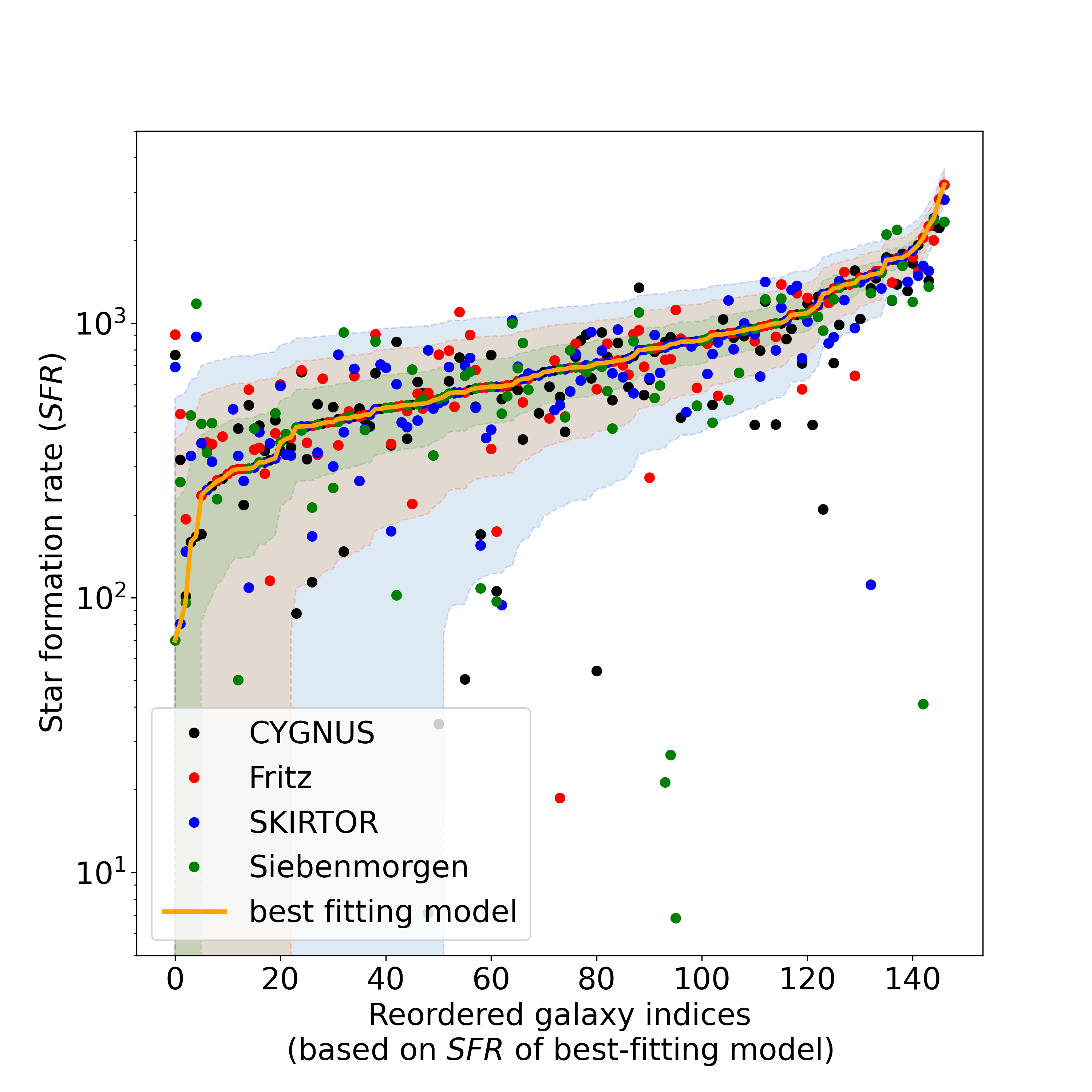}\label{sfr_sph}}
\subfigure[]{\includegraphics[width=0.49\textwidth]{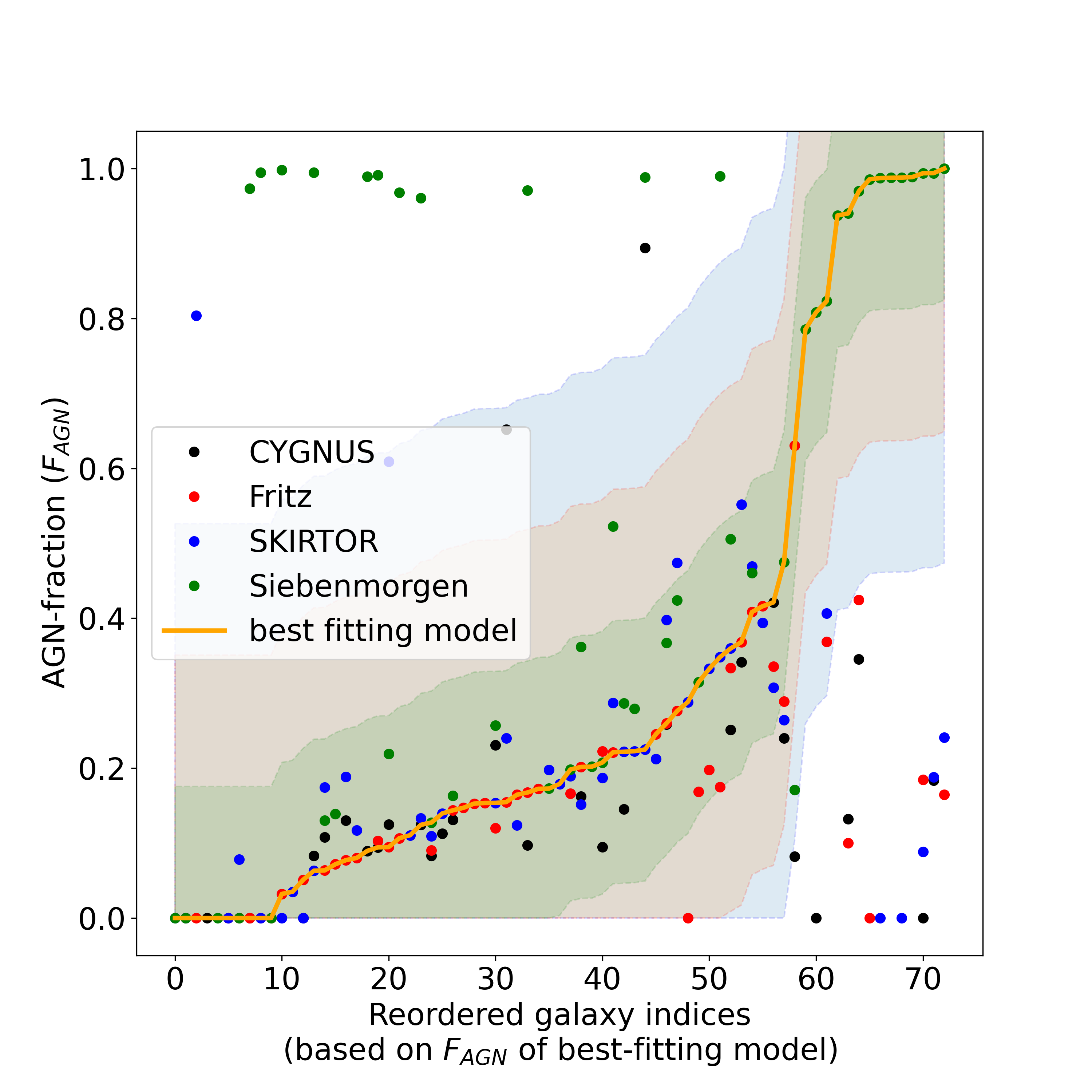}\label{agn_sph}}
\caption{Physical properties of galaxies for the fits that assume a spheroidal geometry for the host. The dots with different colors correspond to different AGN models, while the orange line indicates the properties of the best-fitting model for each galaxy. In each panel, the indexing of the galaxies is done in a way that leads to ascending order for the specific property of the best-fitting model, hence there is no correspondence between panels. The green shaded area indicates the average deviation ($\sigma_{p}^{sph}$) of some property $p$, between the AGN models and the best-fitting model. Accordingly, the red shaded area corresponds to $2\sigma_p^{sph}$ deviation and the blue shaded area to $3\sigma_p^{sph}$ deviation from the best-fitting model. The physical properties that we show are (a) the total luminosity ($L_{tot}^c$ in Table \ref{tab:physical_quantities}), corrected for the anisotropic emission of the AGN torus, (b) total stellar mass ($M_{tot}^*$ in Table \ref{tab:physical_quantities}), (c) star formation rate ($\dot M_{tot}$ in Table \ref{tab:physical_quantities}) and (d) AGN fraction ($F_{AGN}$ in Table \ref{tab:physical_quantities}).}
\label{figs_sph}
\end{figure*}

\begin{figure*}
\centering
Physical properties of galaxies (disc geometry host)\par\smallskip
\subfigure[]{\includegraphics[width=0.49\textwidth]{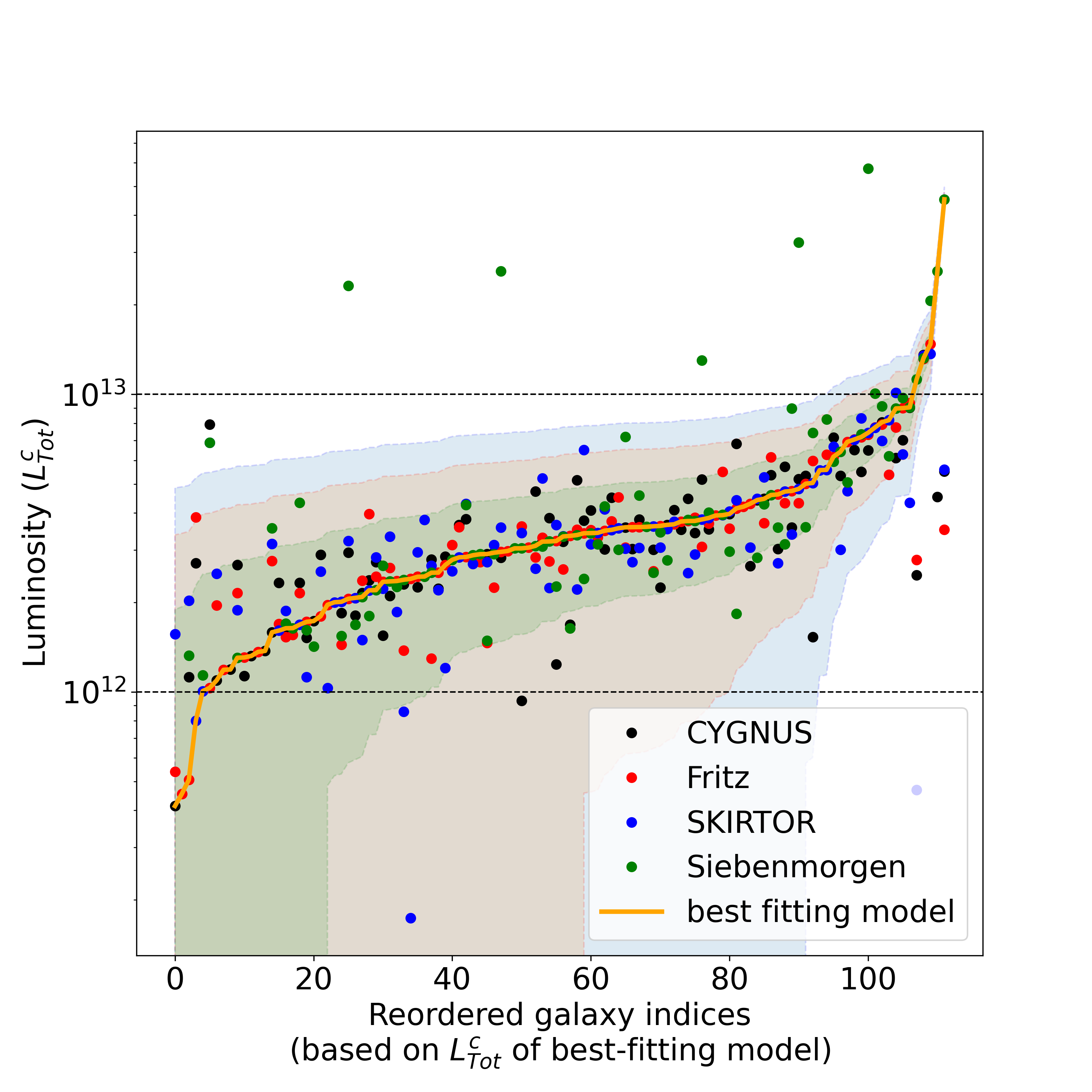}\label{lum_disc}}
\subfigure[]{\includegraphics[width=0.49\textwidth]{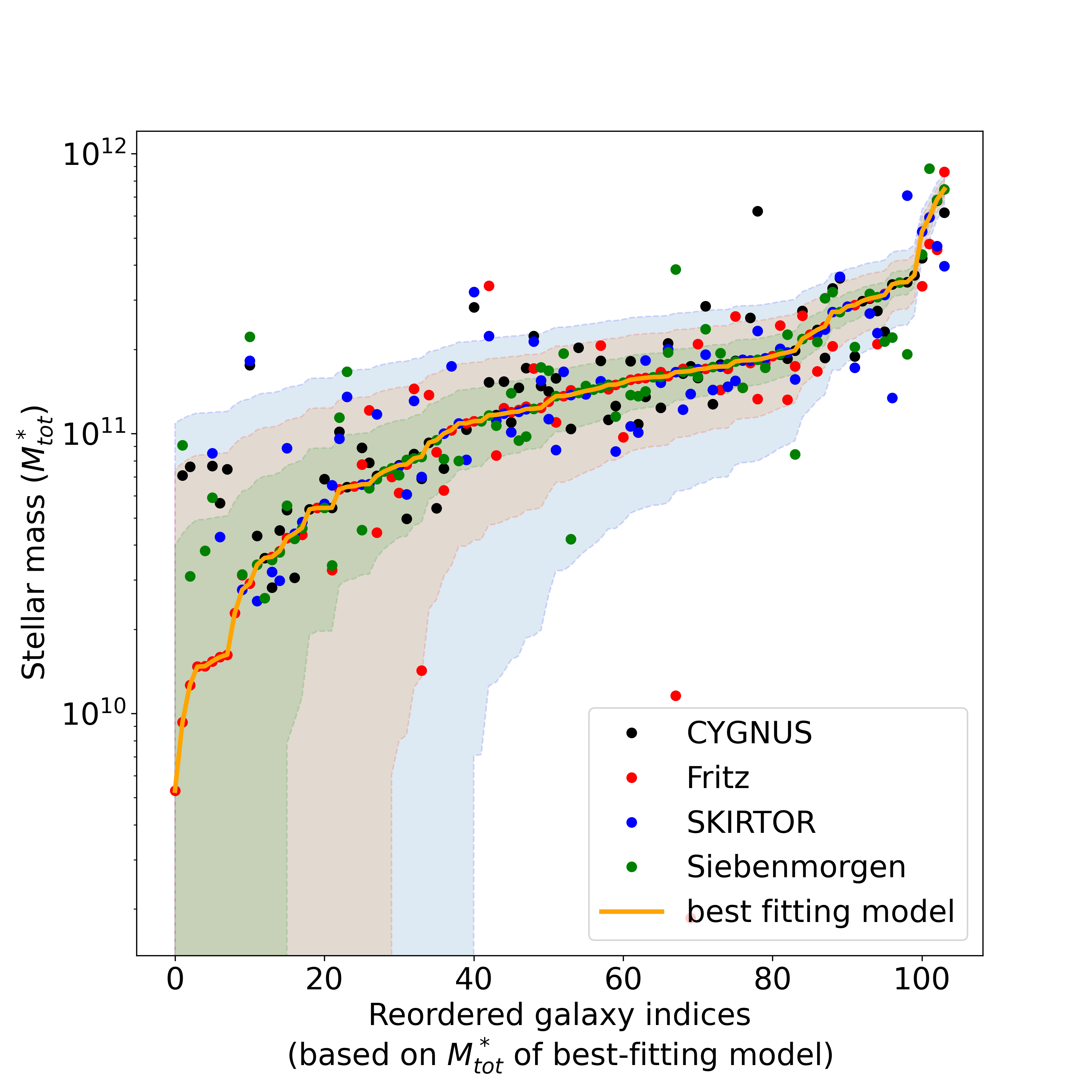}\label{sm_disc}}
\subfigure[]{\includegraphics[width=0.49\textwidth]{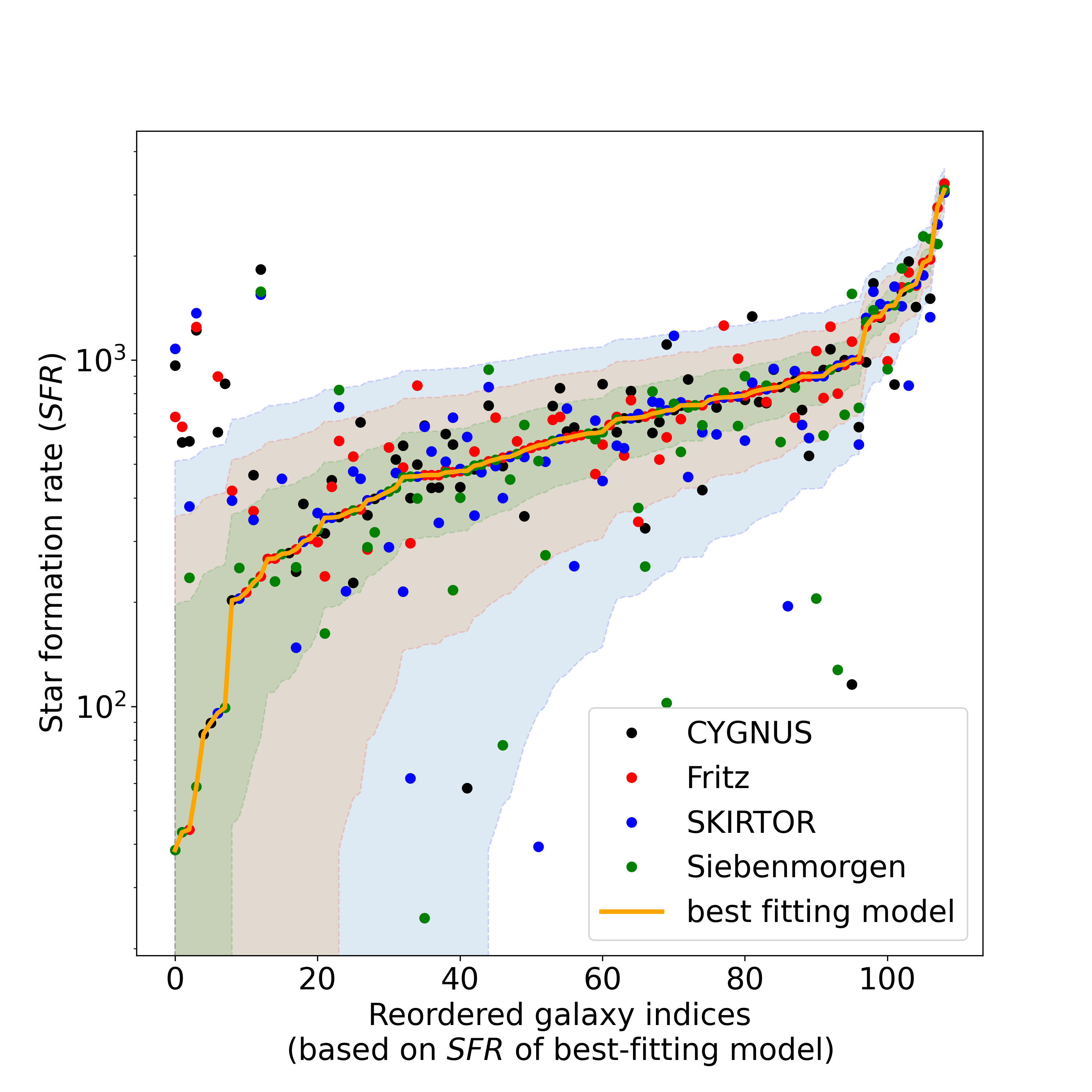}\label{sfr_disc}}
\subfigure[]{\includegraphics[width=0.49\textwidth]{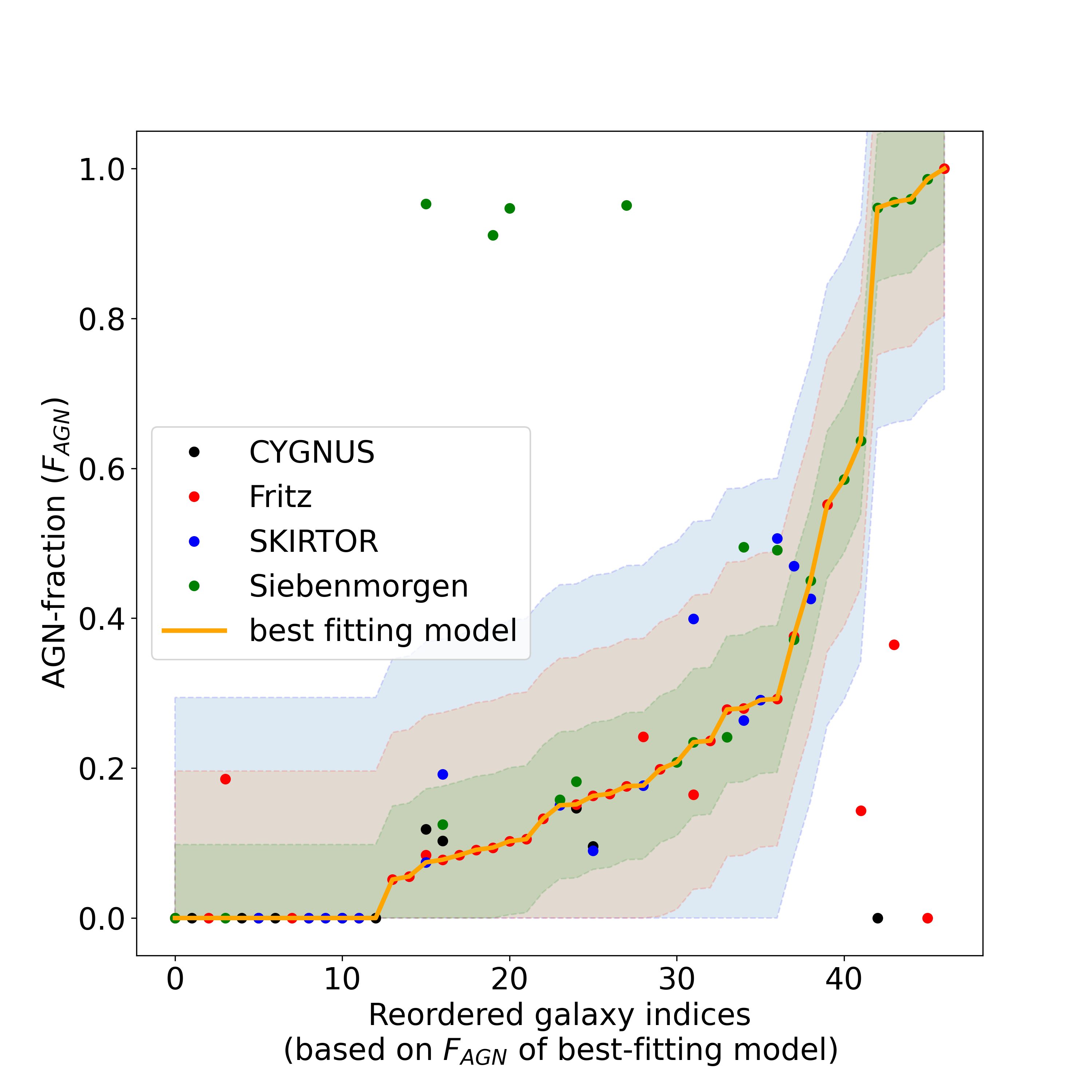}\label{agn_disc}}
\caption{Physical properties of galaxies for the fits that assume a disc geometry for the host. The dots with different colors correspond to different AGN models, while the orange line indicates the properties of the best-fitting model for each galaxy. In each panel, the indexing of the galaxies is done in a way that leads to ascending order for the specific property of the best-fitting model, hence there is no correspondence between panels. The green shaded area indicates the average deviation ($\sigma_{p}^{disc}$) of some property $p$, between the AGN models and the best-fitting model. Accordingly, the red shaded area corresponds to $2\sigma_p^{disc}$ deviation and the blue shaded area to $3\sigma_p^{disc}$ deviation from the best-fitting model. The physical properties that we show are (a) the total luminosity ($L_{tot}^c$ in Table \ref{tab:physical_quantities}), corrected for the anisotropic emission of the AGN torus, (b) total stellar mass ($M_{tot}^*$ in Table \ref{tab:physical_quantities}), (c) star formation rate ($\dot M_{tot}$ in Table \ref{tab:physical_quantities}) and (d) AGN fraction ($F_{AGN}$ in Table \ref{tab:physical_quantities}).}
\label{figs_disc}
\end{figure*}

In this section, we present our findings regarding the main physical properties of our sample (i.e., luminosity, stellar mass, star formation rate, AGN fraction) and try to identify any convergence or difference between models. In all subfigures of Figs.\ref{figs_sph},\ref{figs_disc}, the indexing of the galaxies is done in a way that leads to ascending order for the specific property of the best-fitting model. The indices in different subfigures do not refer to the same galaxies. Moreover, the green shaded area corresponds to the average deviation of all models from the best fit for some parameter $p$, $\sigma_{p} = avg(|best\_fit - models|)$, while the red and blue areas correspond to $2\sigma_{p}$ and $3\sigma_{p}$ respectively. Wherever we mention `correlation', we refer to Pearson correlation.

\begin{figure*}
\centering
\subfigure[]{\includegraphics[width=0.49\textwidth]{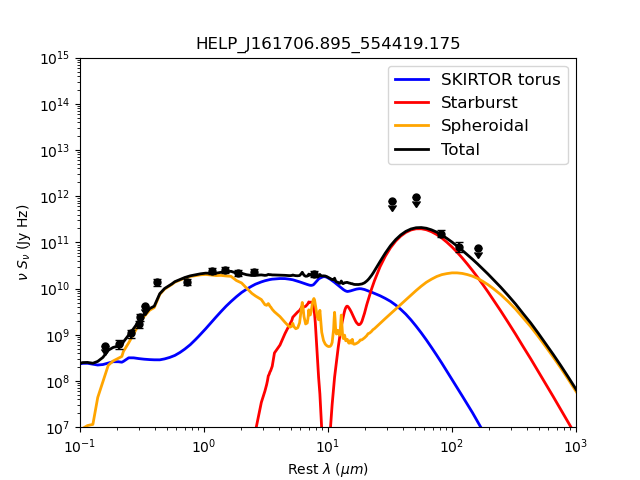}\label{good_fit}}
\subfigure[]{\includegraphics[width=0.49\textwidth]{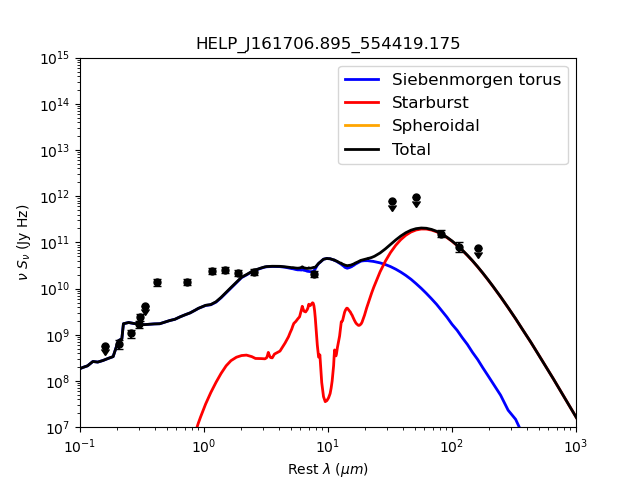}\label{bad_fit}}
\caption{SED fit plots of the galaxy HELP\_J161706.895\_554419.175, with 2 AGN models (SKIRTOR \& Siebenmorgen) utilizing spheroidal geometry for the host. The quality testing for the fits gives $\chi^2_{\nu} = 4.2$ and $\chi^2_{\nu} = 35.9$ for the SKIRTOR and Siebenmorgen models respectively. Obviously, the fit is acceptable when using SKIRTOR but not when choosing the Siebenmorgen model. The upper limits are not used for the calculation of $\chi^2_{\nu}$.
\label{figs_sed-fitting}}
\end{figure*}

\subsection{SED fitting quality}
To begin with, in Fig.\ref{fig_chi_square} we show the quality of the SED fitting process for the spheroidal and disc geometries, in Fig.\ref{chi_sph} and Fig.\ref{chi_disc} respectively. We show the cumulative histograms for the reduced $\chi^2_{\nu}$, with different colors corresponding to different AGN models, as well as the cutoff value $\chi^2_{th} = 5$, shown with the vertical magenta line. The orange curve in both figures shows the cumulative histogram of $\chi^2_{\nu}$ for the `best-fitting model' for each galaxy. Any fit beyond the cutoff point is deemed as a poor fit to the data and is being discarded. For the spheroidal geometry in Fig.\ref{chi_sph}, we see that we can fit roughly $150$ galaxies ($75\%$) with all 4 AGN models. Moreover,  we are able to fit $170$ galaxies (or $85\%$ of the sample) with at least one AGN model. For the disc geometry in Fig.\ref{chi_disc}, we see that the acceptable fits are considerably fewer. We can fit roughly $100$ galaxies ($50\%$) with all of the 4 AGN models, while we manage to fit $121$ galaxies ($60.5\%$) with at least one AGN model. Interestingly, all galaxies that are fitted with a disc host can also be fitted with a spheroidal geometry, utilizing at least one of the AGN models, while the opposite is never true. In Fig.\ref{figs_sed-fitting}, we show an example of the SED fitting results of the galaxy HELP\_J161706.895\_554419.175, with 2 AGN models (SKIRTOR and Siebenmorgen). The SKIRTOR model provides a good fit whereas the Siebenmorgen model gives a bad fit to the data. Particularly, the quality testing for the fits gives $\chi^2_{\nu} = 4.2$ and $\chi^2_{\nu} = 35.9$ for the SKIRTOR and Siebenmorgen models respectively. Obviously, the fit is acceptable when using SKIRTOR but not when choosing the Siebenmorgen model. The data points with an arrow pointing downwards are upper limits and are not taken into consideration when calculating $\chi^2_{\nu}$.

\subsection{Total luminosity}\label{total_luminosity}
In Figs.\ref{lum_sph},\ref{lum_disc} we show the total 1-1000$\mu m$ luminosity ($L^c_{Tot}$ in Table \ref{tab:physical_quantities}), corrected for the anisotropic emission of the AGN torus, for the spheroidal and disc geometries respectively. We show with different colors the various AGN models and with the orange line the best-fitting model for each galaxy. We evaluate the luminosity for $159$ of the galaxies ($79.5\%$) when using the  spheroidal geometry and $112$ of the galaxies ($56\%$) when using the disc geometry. For the spheroidal geometry, the average deviation is $\sigma_L^{sph} \approx 3\times 10^{12}L_\odot$ and for the disc geometry $\sigma_L^{disc} \approx 1.5\times 10^{12}L_\odot$. Although there is variance among the different AGN models, with the bulk being in $\sigma_L$ deviance from the best-fit in both geometries, it is apparent in both figures that our sample is indeed constituted of ULIRGs ($10^{12}L_{\odot} \leq L \leq 10^{13} L_{\odot}$). When using the spheroidal geometry with the Siebenmorgen AGN torus model, about $37$ galaxies ($18.5\%$) have predicted luminosities in the hyperluminous infrared galaxies regime $(L \geq 10^{13}L_{\odot})$. This is probably attributed to the different dust grain size assumption of the Siebenmorgen model.

\begin{figure*}
\centering
\label{sfr_sm}
\subfigure[]{\includegraphics[width=0.49\textwidth]{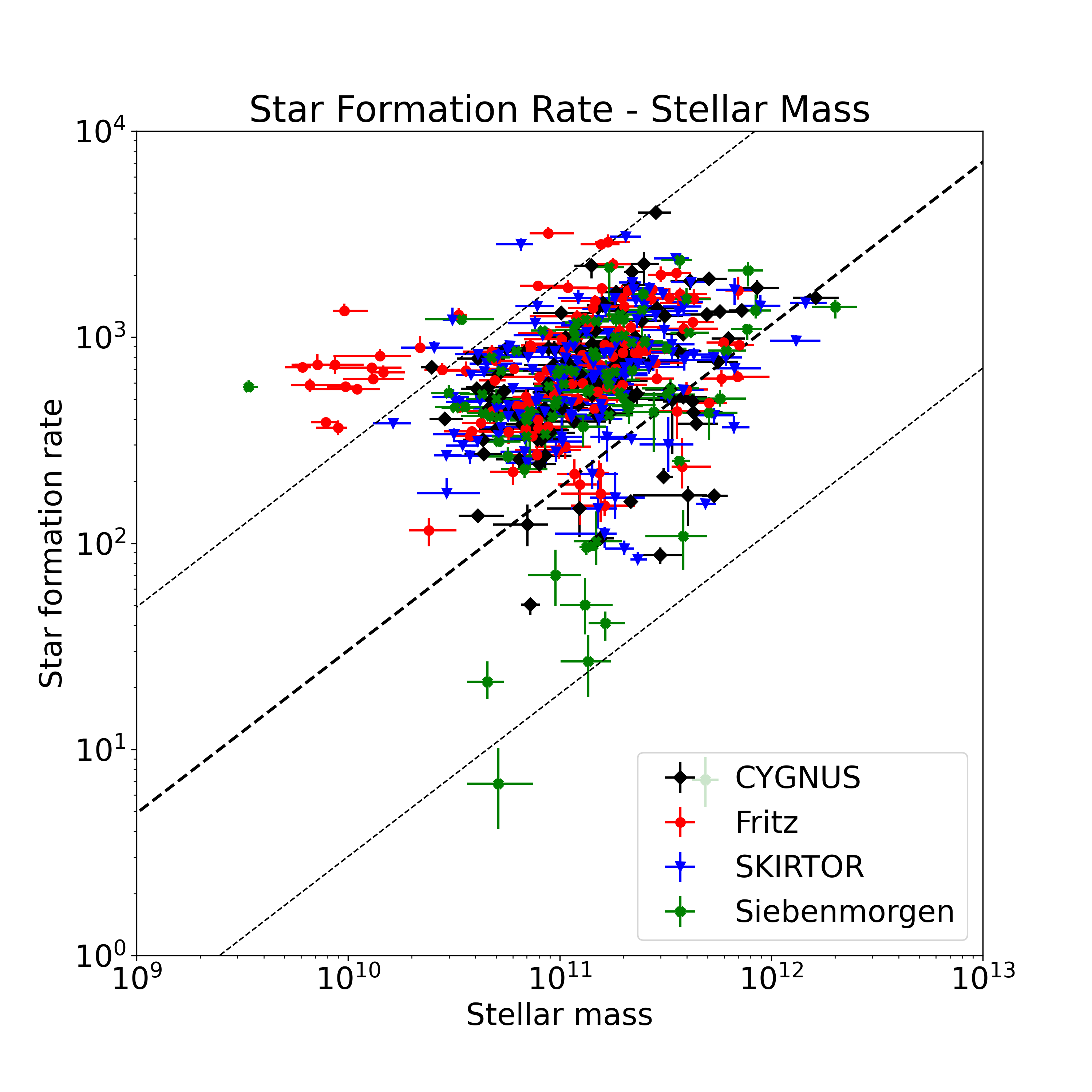}\label{sfr_sm_sph}}
\subfigure[]{\includegraphics[width=0.49\textwidth]{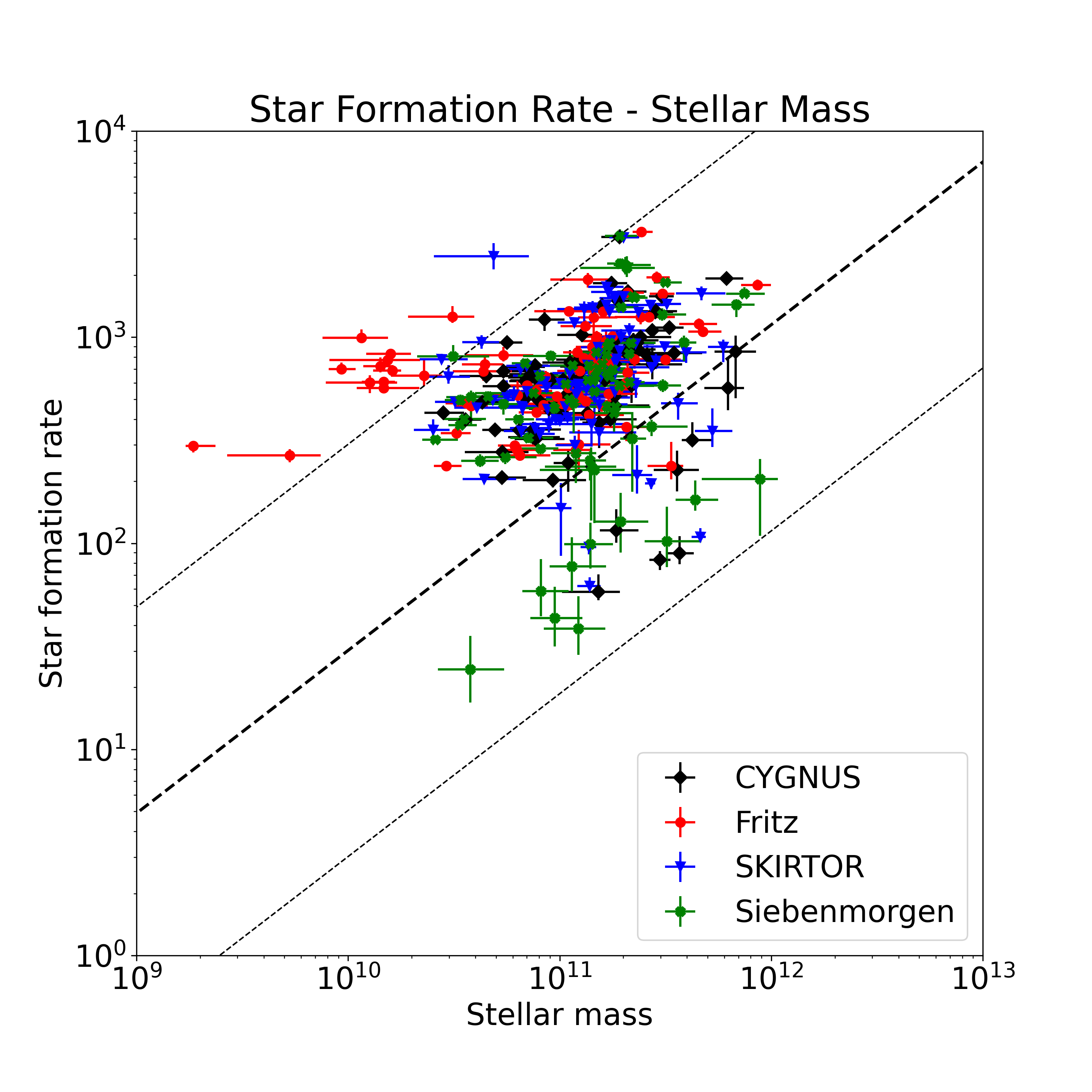}\label{sfr_sm_disc}}
\subfigure[]{\includegraphics[width=0.49\textwidth]{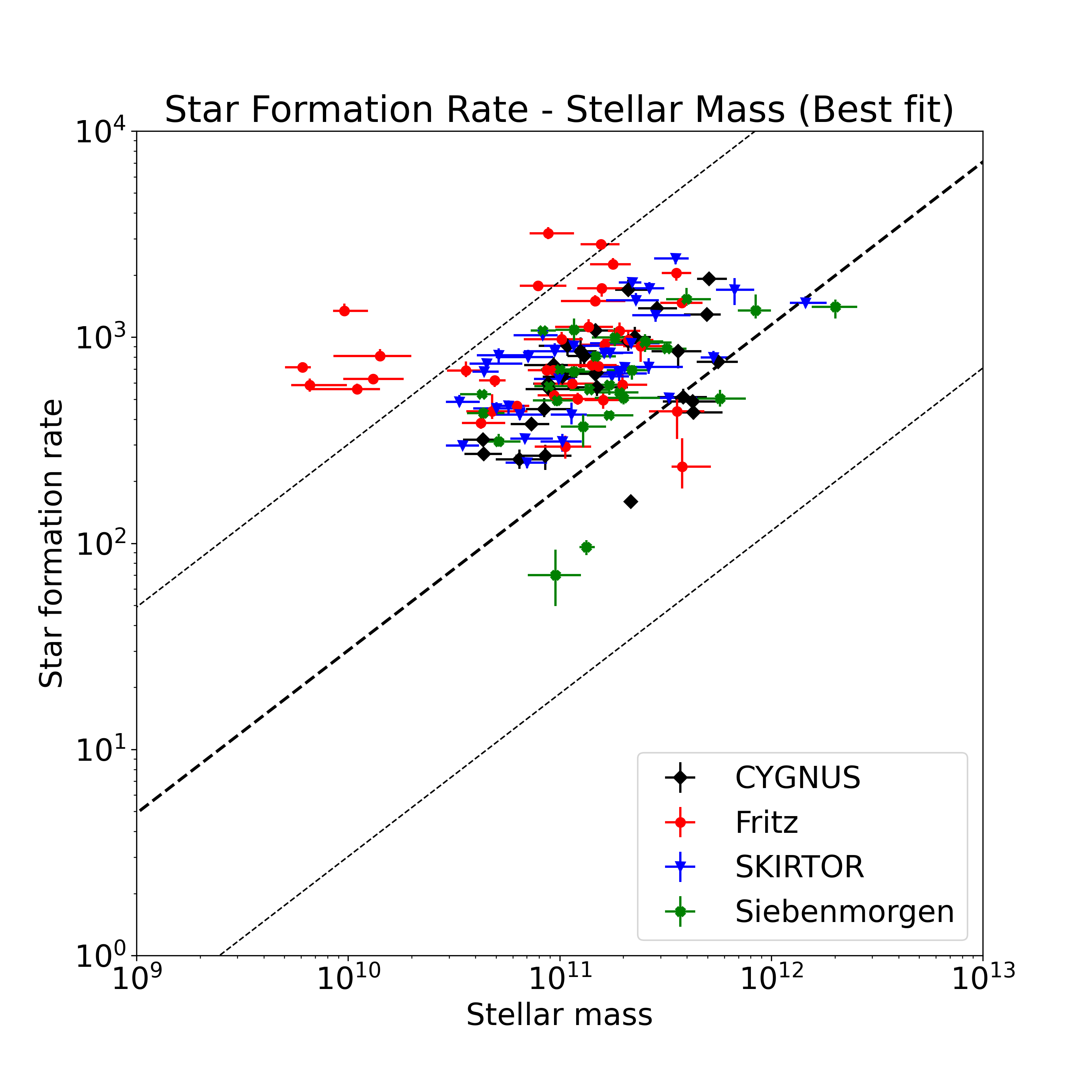}\label{sfr_sm_sph_best}}
\subfigure[]{\includegraphics[width=0.49\textwidth]{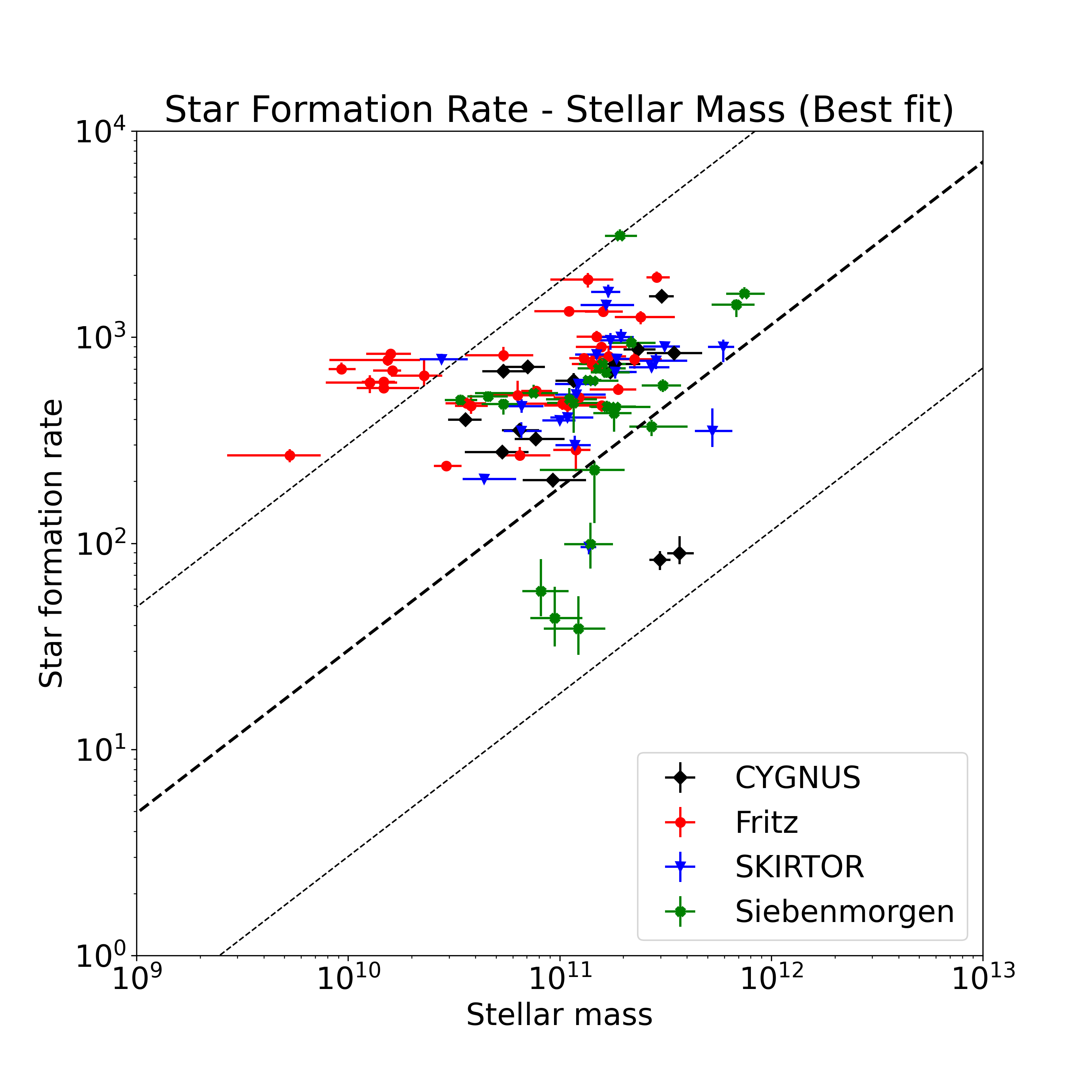}\label{sfr_sm_disc_best}}
\caption{The star formation rate relative to stellar mass for spheroidal geometry (a, c) and disc geometry (b,d). The dashed lines show the galaxy main sequence and the dotted lines deviations from the main sequence by an order of magnitude. We assume $\log(SFR) = a \log(SM) + b$, where $a=0.79$ and $b=-6.42$ were taken from Table 3 of \protect\cite{Speagle_2014}. These figures were inspired and should be compared to \protect\cite{Rodighiero_2011}.}
\end{figure*}

\begin{figure*}
\centering
\includegraphics[scale = 0.55]{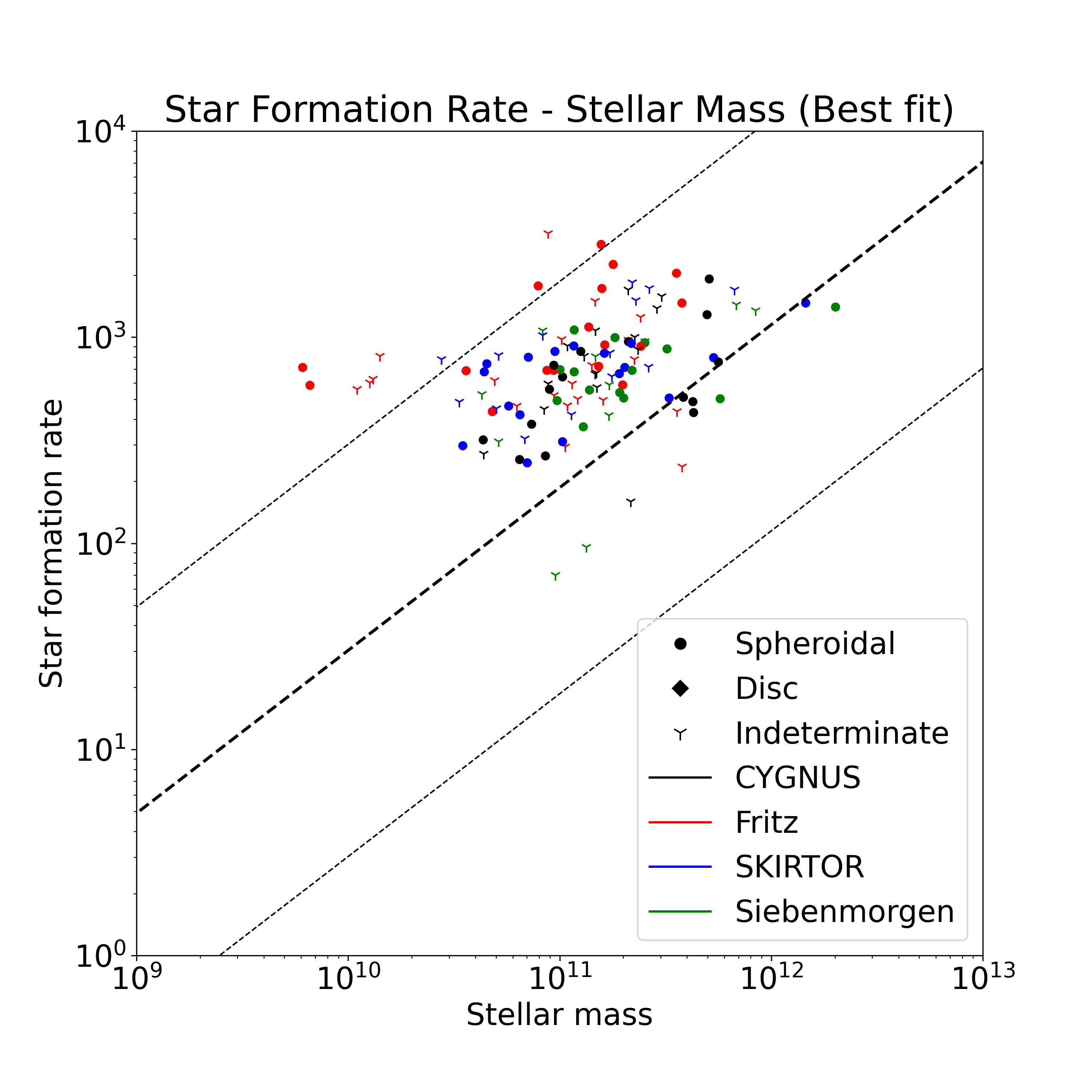}
\caption{The star formation rate relative to stellar mass for the best fit, based on the reduced $\chi^2$ (see text). The dashed lines show the main sequence galaxies and the dotted lines deviations from the main sequence by an order of magnitude. We assume $\log(SFR) = a \log(SM) + b$, where $a=0.79$ and $b=-6.42$ were taken from Table 3 of \protect\cite{Speagle_2014}. This figure was inspired and should be compared to \protect\cite{Rodighiero_2011}.}
\label{sfr_sm_total}
\end{figure*}

\subsection{Stellar mass}
The stellar mass is the physical property for which we observe the best agreement between AGN models. This is reasonable as the stellar mass does not depend very much on the mid-infrared data where the AGN dominates. In Fig.\ref{sm_sph},\ref{sm_disc}, we show the total stellar mass ($M_{tot}^*$ in Table \ref{tab:physical_quantities}) of our sample of galaxies. The different colors correspond to the various AGN models and the orange line shows the stellar mass of the best-fitting model for each galaxy. We evaluate stellar mass for $138$ ($59\%$) galaxies using the spheroidal geometry and $104$ ($52\%$) galaxies when using the disc geometry. The average deviation for the spheroidal geometry is $\sigma_{SM}^{sph} \approx 3.7\times 10^{10} M_\odot$, while for the disc geometry $\sigma_{SM}^{disc} \approx 3.4\times 10^{10} M_\odot$. With an average value of $M_{tot}^* \approx 2\times 10^{11}M_\odot$ and $M_{tot}^* \approx 1.5\times 10^{11}M_\odot$ for our sample, using spheroidal and disc geometries respectively, the average deviance from the best-fit is more than half an order of magnitude smaller, for any AGN model employed. In both figures, the bulk of the galaxies lie within 1 $\sigma_{SM}$ deviation from the best-fit model, while there is also strong correlation between the AGN models and the best-fitting model. In particular, the lowest correlation between any of the 4 AGN models with the best-fitting one is as high as $0.81$ (Fritz) for the spheroidal geometry and $0.80$ (SKIRTOR) for the disc geometry.

\subsection{Star formation rate}
For the star formation rate we similarly see very good agreement between the models as again it does not depend very much on the mid-infrared data . In Fig.\ref{sfr_sph} and Fig.\ref{sfr_disc}, we show the star formation rate ($\dot M_{tot}$ in Table \ref{tab:physical_quantities}) predicted for spheroidal and disc geometries respectively. The different colors correspond to the various AGN models and the orange line shows the star formation rate of the best-fitting model for each galaxy. We evaluate the star formation rate for $147$ ($73.5\%$) galaxies using the spheroidal geometry and $109$ ($54.5\%$) galaxies for the disc geometry. The average deviation for star formation rate is $\sigma_{SFR}^{sph} \approx \sigma_{SFR}^{disc} \approx 150 M_\odot/yr$ for both geometries. The average value of star formation rate for our sample is $\dot M_{tot}^{sph} \approx 800 M_\odot/yr$ and $\dot M_{tot}^{disc} \approx 670 M_\odot/yr$ for the spheroidal and disc geometries respectively. In both figures, the bulk of the galaxies appear to be within $2\sigma_{SFR}$ from the best-fit model. As in the case of the stellar mass, we observe a high correlation of all AGN models with the best-fitting model. In particular, the lowest correlation is as high as $0.76$ (Siebenmorgen and best-fitting model) for the spheroidal geometry and $0.73$ (CYGNUS and best-fitting model) for the disc geometry.

\subsection{AGN fraction}
The last physical property we consider is that of the AGN fraction which we define as the ratio of the corrected AGN luminosity to the total corrected luminosity. An interesting question we explored is whether an AGN is needed at all to explain the SEDs of our sample. To test this we fitted all of the galaxies in the sample with the CYGNUS model assuming the spheroidal geometry, which is usually the best fit, with the AGN component completely switched off. By comparing the reduced $\chi^2$ of the best-fitting combination that includes an AGN and the fit without an AGN in Fig. \ref{fig_no-agn} in the Appendix we see that the fit without an AGN nearly always produces a worse fit. We see this as evidence for the presence of AGN at some level in all of the galaxies in our sample.

In Fig.\ref{agn_sph} and Fig.\ref{agn_disc}, we show the AGN fraction ($F_{AGN}$ in Table \ref{tab:physical_quantities}) for spheroidal and disc geometries respectively. The different colors correspond to the various AGN models and the orange line shows the AGN fraction of the best-fitting model for each galaxy. We can only evaluate the AGN fraction for $73$ ($36.5\%$) galaxies using the spheroidal geometry and $47$ ($23.5\%$) using the disc geometry. The average deviation from the best-fitting model is $\sigma_{AGN}^{sph} \approx 0.170$ and $\sigma_{AGN}^{disc} \approx 0.098$, for the spheroidal and disc geometries respectively. The average value of AGN fraction in our sample is $F_{AGN} \approx 0.32$ and $F_{AGN} \approx 0.24$ for the spheroidal and disc geometries respectively. Evidently, the subfigures for AGN fraction (Fig.\ref{agn_sph} and Fig.\ref{agn_disc}) are a lot sparser compared to the other properties. This is due to the fact that $F_{AGN}$ is not well-constrained (see paragraph before Section \ref{sec_best_fit}) from the SED fitting, especially when using the disc geometry. The correlation values here range from $0$ to $0.9$, depending on the AGN model and the geometry used. However, the statistical significance of these values is low, as they are derived from small vectors containing many NaN values. In terms of agreement between models, it appears that all models are within $\sigma_{AGN}$ for both geometries, with the occasional exception of the Siebenmorgen model. Especially when using the spheroidal geometry, the Siebenmorgen model consistently overestimates the AGN fraction, possibly for the same reason discussed in Section \ref{total_luminosity}.

Moreover, we look at the AGN fraction computed without the anisotropy correction for the emission of the torus. We define the raw AGN fraction as the ratio of the raw AGN luminosity to the total raw luminosity. We do this in order to examine whether the differences that we find are due to the anisotropy correction that we implement or due to the different assumptions of the models for the AGN torus. In Fig.\ref{fig_anisotropic_agn} in Appendix \ref{appendix}, we show the equivalent figures to Figs.\ref{agn_sph},\ref{agn_disc}. In these figures we see a similar picture as in the figures that include the anisotropy correction. We therefore conclude that the differences in the AGN fraction are mainly due to the models themselves and not due to the applied anisotropy correction for the torus emission.

\subsection{Main sequence}
It is evident that we are able to evaluate stellar mass and star formation rate for a considerable percentage of galaxies from our sample. Moreover, when we do evaluate these properties, the high correlation values indicate that there exists little to no dependence on the adopted AGN torus model. This comes from the fact that stellar mass and star formation rate are mainly determined by the UV-near-infrared data as well as the far-infrared data, where there is generally very little contribution from the AGN torus. The good estimates on stellar mass and star formation rate allow us to investigate the position of our sample with respect to main sequence galaxies. Inspired by \cite{Rodighiero_2011}, we plot the star formation rate versus the stellar mass in Fig.\ref{sfr_sm}. In all subfigures, we show the main sequence (bold dashed line), derived from \cite{Speagle_2014} for $z=2$, as well as a factor of 10 deviation from it. The radiative transfer models that we use and the derived main sequence lines assume the same initial mass function (IMF), namely the Salpeter IMF. For both spheroidal and disc geometries, shown in Figs.\ref{sfr_sm_sph},\ref{sfr_sm_disc} respectively, the bulk of the galaxies lie just above main sequence, indicating high star formation rate, with some exceptions. Retaining just the best fitting AGN model in both cases, we show for the spheroidal geometry in Fig.\ref{sfr_sm_sph_best}, 126 galaxies whose stellar mass and star formation rate is constrained and for the disc geometry in Fig.\ref{sfr_sm_disc_best}, 96 galaxies.

\subsection{Geometry of the host}
Now, we explore whether it is possible to discern the geometry of the galaxies in our sample from SED fitting. For this purpose, and for each galaxy, we compare the reduced $\chi_{\nu}^2$ value of the best fitting spheroidal and disc geometry fits. If the difference in $\chi_{\nu}^2$ is larger than $1$, we identify the galaxy to have the geometry with the minimum $\chi_{\nu}^2$. If the difference is smaller than $1$, we refer to the galaxy as of indeterminate geometry. We show this in Fig.\ref{sfr_sm_total}, where we see that 63 galaxies $(31.5\%)$ are of spheroidal geometry, 61 galaxies $(30.5\%)$ are of indeterminate geometry and most importantly, there are no galaxies that are identified with a disc geometry host. In this figure we do not show the 30 galaxies $(15\%)$ that have $\chi_{\nu}^2>5$ for all AGN torus models. Finally, there are 46 galaxies $(23\%)$ where the best fitting model does not constrain the stellar mass and star formation rate and are therefore not included in the figure.

\subsection{Comparison with energy balance methods (CIGALE)}
\begin{figure*}
\label{cigale}
\centering
\subfigure[]{\includegraphics[width=0.49\textwidth]{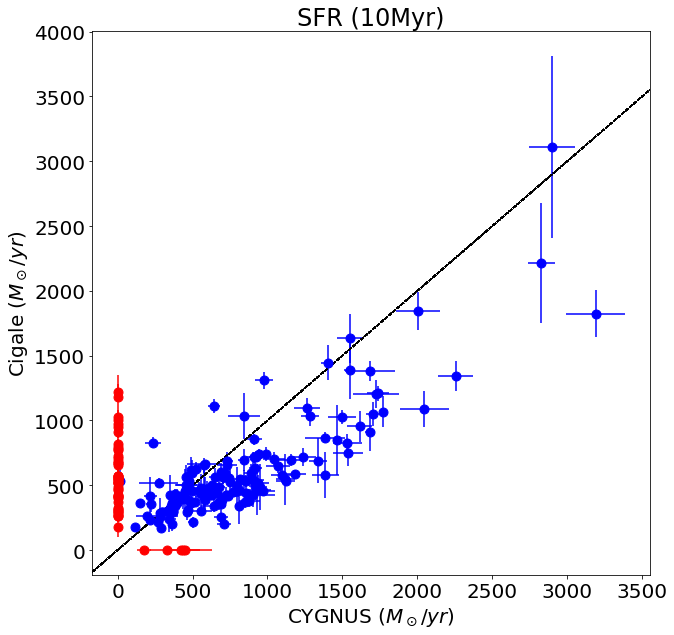}\label{cigale_sfr10}}
\subfigure[]{\includegraphics[width=0.49\textwidth]{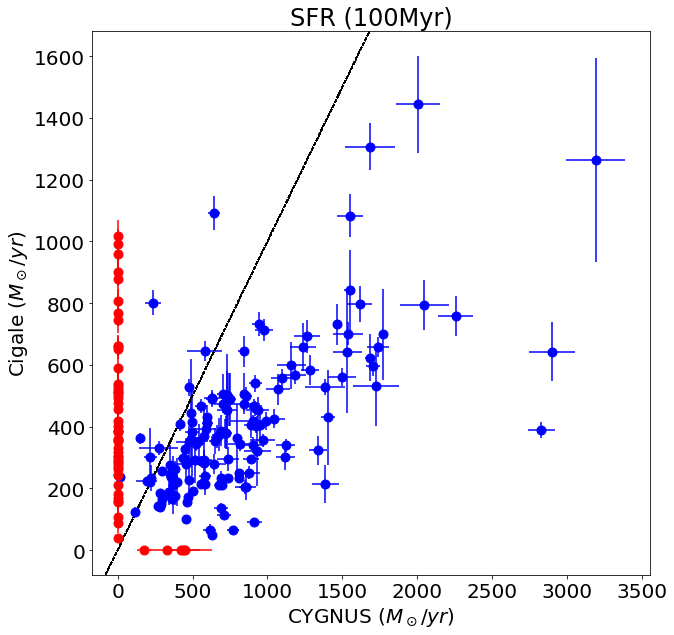}\label{cigale_sfr100}}
\subfigure[]{\includegraphics[width=0.49\textwidth]{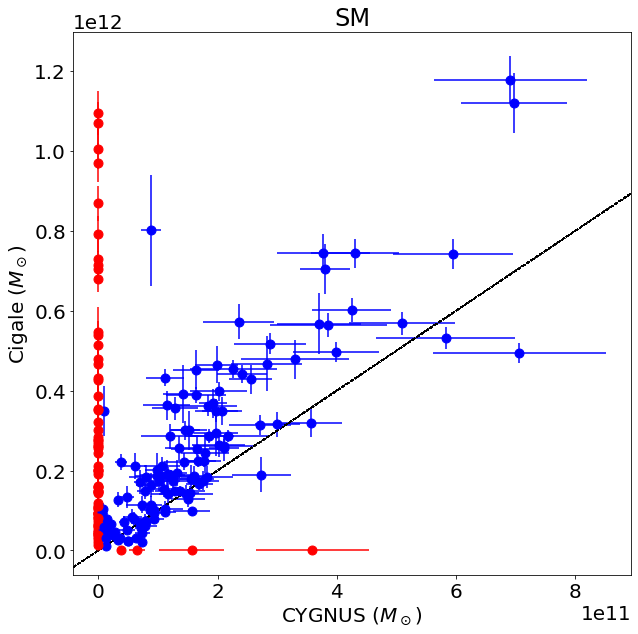}\label{cigale_sm}}
\subfigure[]{\includegraphics[width=0.49\textwidth]{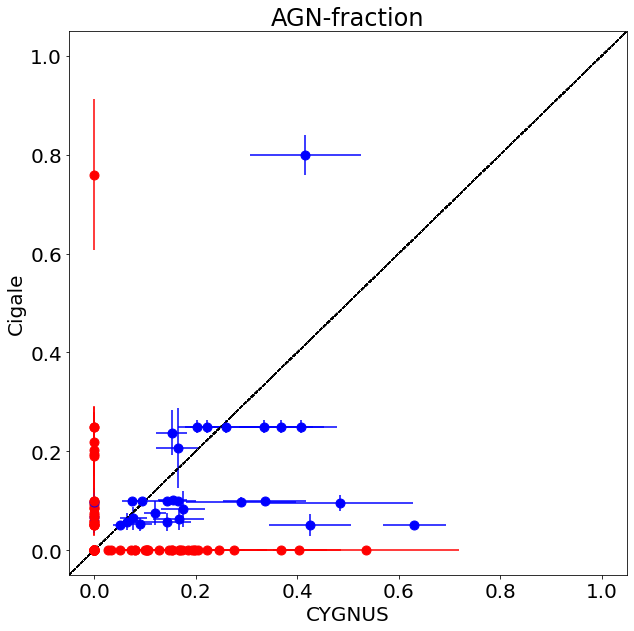}\label{cigale_agn}}
\caption{Comparison of the key physical properties of galaxies for the CYGNUS fitting (performed in this work) and the CIGALE fitting (performed in \protect\cite{malek2018}). The blue dots refer to galaxies that the corresponding property is constrained for both fitting methods, while the red dots refer to galaxies where one of the fitting methods fails to constrain the property. In (a) we show the comparison for the SFR estimates for the CYGNUS and CIGALE fittings. For the CIGALE output, SFR is averaged over $10$Myr. Accordingly, in (b) we show the comparison for the SFR estimates for the CYGNUS and CIGALE fittings, when averaged over $100$Myr. In (c) we show the comparison for SM and in (d) we show the comparison for AGN fraction. }
\end{figure*}

Finally, we perform a brief comparison with the output obtained in \cite{malek2018} using CIGALE. A large part of our sample, $193$ out of the $200$ galaxies, is fitted in \cite{malek2018}. In that work, the authors choose the Chabrier IMF, making it necessary to convert their output before comparing with our results, since we use the Salpeter IMF. We therefore divided their SFR estimates with $0.63$ and their SM estimates with $0.61$. Moreover, the authors in \cite{malek2018} assume the Fritz model for the AGN torus. Consequently, we will use the output of the Fritz model in our fittings as well, in order to compare like with like. As mentioned in Section \ref{sec_fitting}, in order to accept a physical property as `well constrained', we require that the value of the property is larger than 2$\sigma$, where $\sigma$ the error of the property. In Fig.\ref{cigale} we show how the key physical quantities compare between the two fitting methods. For all subfigures, the blue points refer to galaxies for which the corresponding property is well constrained for both fitting methods. The red points refer to galaxies for which one of the fitting methods fails to satisfy $2\sigma$ accuracy and we set that value to zero. 

In Figs.\ref{cigale_sfr10},\ref{cigale_sfr100} we show the comparison for the estimates of SFR for CYGNUS and CIGALE, averaged over $10$ Myr and $100$ Myr respectively. When using CYGNUS, with the Fritz model for the torus, we constrain SFR for $149$ galaxies, while for CIGALE we have both estimates of SFR for $192$ galaxies. We see that there is good general agreement between the estimates of SFR with CYGNUS and CIGALE, when averaged over $10$ Myr (Fig.\ref{cigale_sfr10}). Our results tend to estimate a slightly higher SFR, which could be due to the fact that our method attributes more of the far-infrared emission to obscured star formation. When averaging the SFR in CIGALE over $100$ Myr (Fig.\ref{cigale_sfr100}), the difference of the two estimates increases, which is to be expected. In Fig.\ref{cigale_sm} we show the comparison for the SM estimates. Here, we constrain SM for $149$ galaxies when using CYGNUS and for $193$ galaxies with CIGALE. We also find good general agreement between the two estimates, with a slight overestimation in CIGALE. This can be attributed to the same reason we see an underestimation in SFR. Finally, in Fig.\ref{cigale_agn} we show the comparison for the AGN fraction, constrained for $76$ galaxies when using CYGNUS and for $52$ galaxies for CIGALE. Here we see that there is some general agreement in the regime below $0.2$, but the sample is too small to make a proper comparison between the two fitting methods. Some of the difference may be due to the fact that CIGALE does not apply anisotropy corrections for the AGN luminosity.

\section{Discussion and Conclusions}\label{sec_discussion}

We studied a sample of 200 galaxies at $z\approx 2$ selected from the ELAIS N1 field which have optical to submillimeter data provided by the HELP project. We fitted our sample with the CYGNUS multi-component radiative transfer models using an MCMC code. We use four different models for the AGN torus component in an attempt to quantify the uncertainties introduced by the choice of the particular model for this component. Our main conclusions are the following

\begin{enumerate}

\item We find that there is good agreement in the estimates of SFR and stellar mass by all combinations of models.

\item We have demonstrated that there is big uncertainty in the predicted AGN fraction. One reason for this is that we take properly into account the anisotropy of the emission of the torus as predicted by the models themselves. The main reason though is the big variety of AGN torus models we use which explore variation of the geometry (tapered/flared/full sphere), nature of the dusty medium (smooth/two-phase), dust composition (normal dust/fluffy dust), presence of polar dust \citep{siebenmorgen2015} or not. This highlights the importance of knowing which model best describes AGN tori. This is of course an active field of research.

\item We find that the \cite{siebenmorgen2015} model predicts very high AGN luminosities for some galaxies. This is most probably related to the fact that this model assumes fluffy grains which have large emissivity in the far-infrared. With this model it is therefore possible to find solutions where most of the far-infrared emission is attributed to the AGN and not to the starburst or host galaxy. The high AGN luminosities predicted by the \cite{siebenmorgen2015} model are also associated with lower SFRs. If this is the preferred model, this could help relieve tension between observed extreme starbursts with SFRs of the order of thousands of solar masses per year, and galaxy formation models that predict such starbursts should not exist.

\item An important novelty of this work is that we use physically motivated radiative transfer models for all components that model in a realistic way the star-dust geometry. This is in contrast to results using SED codes based on energy balance (e.g. CIGALE). A comparison of our results with those obtained by \cite{malek2018}  shows that CIGALE overestimates the stellar mass and underestimates the SFR. We plan to make a more thorough comparison with a larger sample in future work.

\item An interesting result coming out of this study is that we determined the fraction of galaxies at $z\sim2$ that are best fitted with spheroids and discs.  We find the majority are spheroids which lie above the main sequence. This may be related to our selection which may preferentially pick the most massive galaxies with stellar mass  $> 4 \times 10^{10} M_\odot$. Our result is consistent with recent results from JWST \citep{lee2023} who find that in the high stellar mass regime the dominant morphology is spheroidal. A similar conclusion was reached by \cite{lofaro2015} who compared fits of radiative transfer models of spheroidal and disc galaxies for a sample of massive galaxies at $z \sim 2$. 

\end{enumerate}

\section*{Acknowledgements}

We would like to thank an anonymous referee for useful comments and suggestions. This project has received funding from the European Union’s Horizon 2020 research and innovation programme under the Marie Skłodowska-Curie grant agreement No 859937. 

%%%%%%%%%%%%%%%%%%%%%%%%%%%%%%%%%%%%%%%%%%%%%%%%%%
\section*{Data Availability}

The raw data SED data which are used in this paper are available at the public database of the Herschel Extragalactic Legacy Project (HELP): (\url{https://hedam.lam.fr/HELP/}). The processed data, after SED fitting, are available on demand.



\bibliographystyle{mnras}
\bibliography{example} % if your bibtex file is called example.bib

\appendix\label{appendix}

\section{Complementary Figures}

\begin{figure*}
\centering
Fits with no AGN component\par\smallskip
\includegraphics[width=0.5\textwidth]{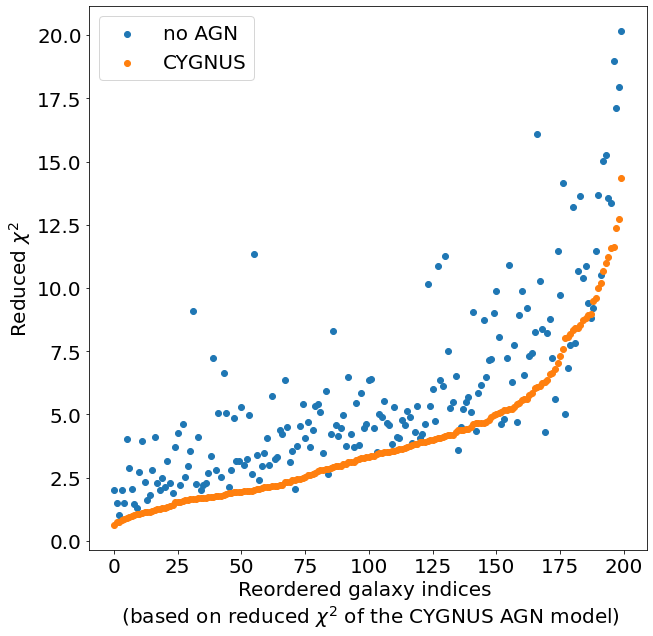}
\caption{Comparison of the reduced $\chi^2$ value of the best fitting AGN model (orange) and a fit with no AGN model included (blue). The galaxy indices are reordered in a way that the reduced $\chi^2$ when using the CYGNUS AGN model is in ascending order.}
\label{fig_no-agn}
\end{figure*}

\begin{figure*}
\centering
Data Sampling\par\smallskip
\includegraphics[width=0.5\textwidth]{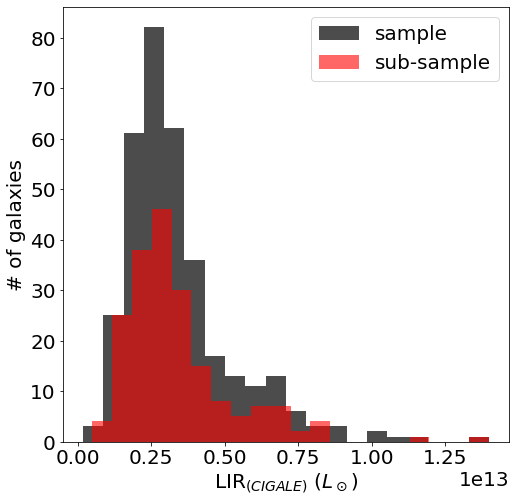}
\caption{The LIR estimates for our $200$ galaxy sub-sample (red), compared to the LIR estimates for $340$ galaxies out of the initial sample of $354$ galaxies (black). These estimates are taken from the CIGALE fitting performed in \protect\cite{malek2018}.}
\label{fig_lir}
\end{figure*}

\begin{figure*}
\centering
Star formation dependence in torus ratio of outer to inner radius\par\smallskip
\includegraphics[width=0.55\textwidth]{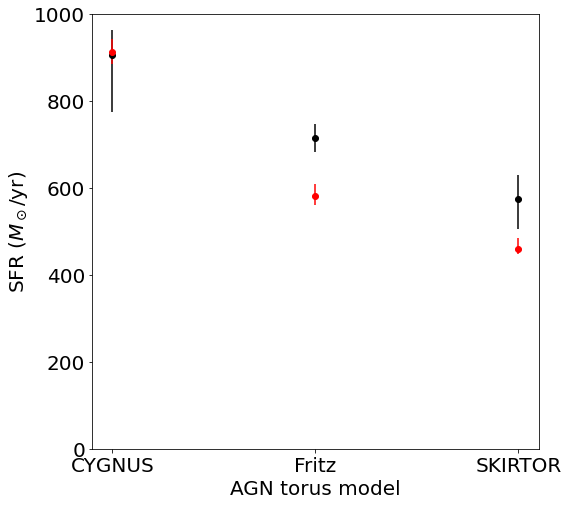}
\caption{In this figure we explore the dependence of SFR estimates on the ratio of outer to inner torus radius for the galaxy $HELP\_J161123.428+535540.996$. The black dots correspond to the value of SFR when the ratio of outer to inner radius is fixed to the values of Table \ref{tab:fiting_parameters} and the red dots correspond to the values of SFR for when the parameter is free. The range of the parameter varies with the AGN model and the ranges are assumed as in \protect\cite{efstathiou2022}.}
\label{fig_torus_radial_extent}
\end{figure*}

\begin{figure*}
\centering
AGN fraction (with no anisotropic correction)\par\smallskip
\subfigure[]{\includegraphics[width=0.49\textwidth]{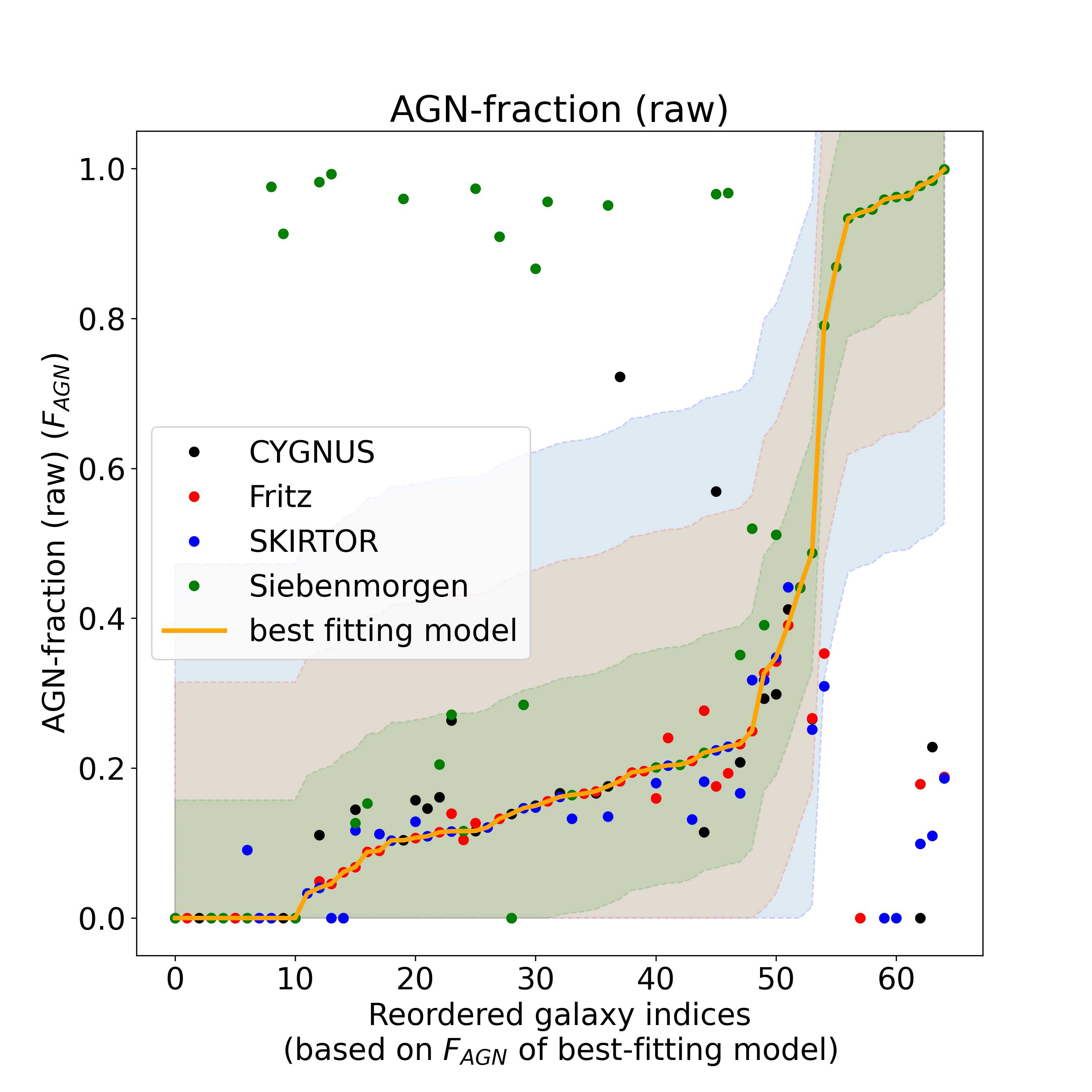}\label{agn_raw_sph}}
\subfigure[]{\includegraphics[width=0.49\textwidth]{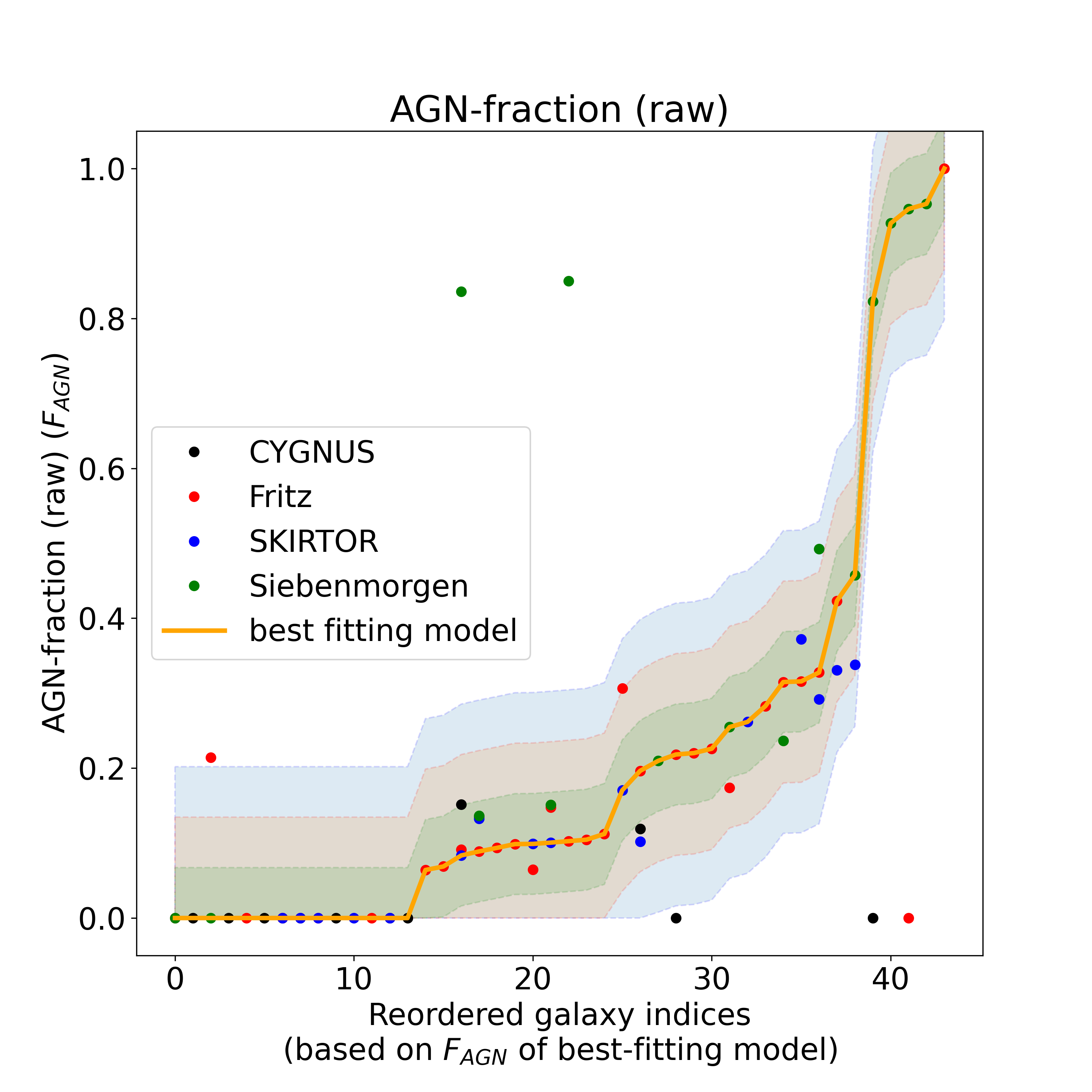}\label{agn_raw_disc}}
\caption{AGN fraction (with no anisotropic correction). The dots with different colors correspond to different AGN models, while the orange line indicates the properties of the best-fitting model for each galaxy. On the left panel we show the raw AGN fraction for the spheroidal host and on the right panel the raw AGN fraction for the disc host. In each panel, the indexing of the galaxies is done in a way that leads to ascending order for the specific property of the best-fitting model, hence there is no correspondence between panels. The green shaded area indicates the average deviation ($\sigma_{p}^{sph}$) of some property $p$, between the AGN models and the best-fitting model. Accordingly, the red shaded area corresponds to $2\sigma_p^{sph}$ deviation and the blue shaded area to $3\sigma_p^{sph}$ deviation from the best-fitting model.}
\label{fig_anisotropic_agn}
\end{figure*}

% Don't change these lines
\bsp	% typesetting comment
\label{lastpage}

\end{document}